\documentclass[preprintnumbers, floatfix, twocolumn, preprintnumbers, iop, superscriptaddress]{revtex4}
\usepackage{amsfonts}
\usepackage{amsmath}
\usepackage{amssymb,epsf}
\usepackage{latexsym}
\usepackage{graphicx,epsfig}
\usepackage{amssymb}
\usepackage{subfigure}
\usepackage[colorlinks=true,citecolor=blue,linkcolor=blue,urlcolor=black]{hyperref}
\usepackage{epstopdf}
\usepackage{color}
\usepackage{enumitem}

\def\be{\begin{equation}}
\def\ee{\end{equation}}
\def\ba{\begin{eqnarray}}
\def\ea{\end{eqnarray}}

\def\ep{\E}
\def\E{\varepsilon}

\begin{document}

\title{Magnetic Interactions and Cluster Formation: Boosting Surface Thermopower in Topological Insulators}
\author{M. Tirgar}
\author{H. Barati Abgarmi}
\author{J. Abouie}
\email{jahan@iasbs.ac.ir}
\affiliation{Department of Physics, Institute for Advanced Studies in Basic Sciences (IASBS), Zanjan 45137-66731, Iran}

\begin{abstract}	
This study presents a theoretical investigation of the thermoelectric properties of three-dimensional magnetic topological insulators (TIs), with a focus on the role of exchange interactions between magnetic dopants. The presence of these magnetic atoms on the TI surface modulates the local magnetic order, which in turn alters the electronic band structure and surface transport phenomena. Magnetic correlations, such as those arising from ferromagnetic or antiferromagnetic exchange, promote cluster formation, magnetic domain structures, and spin fluctuations, all of which critically influence thermoelectric responses. Using extensive Monte Carlo simulations based on Ising and Heisenberg models of these surface exchange interactions, we analyze how magnetic clustering, particularly near the surface critical temperature, affects relaxation dynamics, electrical and thermal resistivity, the Seebeck coefficient, and the thermoelectric figure of merit. Our results demonstrate that exchange-driven magnetic clustering enhances the scattering of Dirac surface states, thereby increasing the thermoelectric power factor. Specifically, optimized interlayer and intralayer exchange interactions can elevate the surface thermopower beyond levels observed in conventional spin-based thermoelectric materials. These findings highlight the significant potential of magnetic TIs for thermoelectric applications and provide a foundation for future experimental and theoretical studies of magnetic correlations in topologically nontrivial systems.

\end{abstract}

\maketitle

\section{Introduction}\label{Intro}

Achieving net-zero greenhouse gas emissions is crucial in our lives today, and developing systems capable of generating clean energy is a vital task across various research fields\cite{Fenwick2020}. Thermoelectric generators, which convert waste heat into clean energy, are as efficient as other renewable energy sources like solar cells\cite{Gayner2016}. The pursuit of high-performance thermoelectric materials has garnered significant attention due to their potential applications in technology\cite{Singh2024}. The primary challenge in developing these materials lies in the strong interdependence of the thermopower factor, electrical conductivity, and thermal conductivity. Typically, increasing electrical conductivity decreases thermopower and increases thermal conductivity. Therefore, strategies for developing new thermoelectric materials should focus on mitigating this interdependence, requiring innovative approaches and the use of materials with unique properties\cite{Ivanov2018}.

While thermoelectric coefficients are generally interdependent in conventional materials, magnetic materials with spin-driven thermoelectrics can significantly alter this relationship. For instance, in anisotropic layered magnetic systems such as antiferromagnetic MnTe, thermopower displays anomalous enhancement in the disordered paramagnetic phase above the Neel temperature\cite{Zheng2019}. In this system, the paramagnon-electron drag thermopower has emerged as a compelling phenomenon, exhibiting continuous enhancement in the paramagnetic phase and offering the potential for high-performance thermoelectric devices\cite{Polash2020, Polash2021, Heydarinasab2024}.

Topological systems are also excellent thermoelectric materials, exhibiting strong spin-orbit coupling crucial for spin-driven thermoelectrics\cite{Xu2017, Pan2025}. Topological insulators (TIs) exhibit insulating behavior in their bulk while possessing conductive surface states \cite{Moore2010}. Surface electrons with a Dirac-like spectrum are resistant to perturbations that preserve the system's time-reversal symmetry. When doped with magnetic atoms, these surface states develop an energy gap, leading to exotic phenomena such as the anomalous Hall effect \cite{Tokura2019} and giant anisotropic magnetoresistance\cite{Vyborny2009, Sabzali2015}. In TIs lightly doped with magnetic atoms, the anisotropic relaxation time of surface electrons results in anisotropic magnetoresistance. 

In magnetic TIs such as the van der Waals TI $\rm {MnBi_2Te_4}$, interactions of magnetic atoms and the sequential scattering of surface electrons significantly influence transport properties. These magnetic atoms form temperature-dependent clusters on the surface, affecting the material's magnetic, electric, and thermoelectric characteristics\cite{Zarezad2018, Zarezad2020}.

In this paper, we investigate the thermoelectric properties of the surface of intrinsic magnetic topological insulators (TIs) by examining the effects of exchange interactions among magnetic atoms. On the surface of magnetic TIs, these atoms typically reside on a triangular lattice and interact with their neighbors via an exchange spin Hamiltonian, leading to the formation of magnetic clusters. The formation of these clusters through exchange interactions increases the scattering of Dirac electrons, thereby boosting thermoelectric efficiency. The clusters are randomly distributed across the surface of the TIs and act as scattering centers, influencing surface transport properties.

We apply the semiclassical Boltzmann formalism to calculate relaxation times, electrical and thermal resistivities, the thermopower factor, and thermoelectric figure of merit as functions of the mean size and number of clusters. Extensive Monte Carlo (MC) simulations \cite{diep2011} are performed to obtain the temperature dependence of the mean size and number of clusters. We compare magnetic clustering on the surface of TIs where the exchange interaction between magnetic atoms is modeled by Ising and Heisenberg models, focusing on cluster definitions relative to surface critical temperatures. Our simulations encompass various configurations, including ferromagnetic and antiferromagnetic bulk phases.

Our analysis reveals that the size and number of magnetic clusters significantly influence the relaxation dynamics, electrical and thermal resistivity, thermopower, and the thermoelectric figure of merit. We demonstrate that optimized interlayer and intralayer exchange interactions, along with specific Fermi energies, can elevate the surface thermopower of TIs to approximately 200 $\mu$V/K. This value is comparable to that of high-performance thermoelectrics like $\text{Bi}_2\text{Te}_3$ and its alloys, as well as antiferromagnetic MnTe, a material used in spin-based thermoelectrics where paramagnon drag induces anomalously high thermopower above the Néel temperature (typically ranging from 100 to 200 $\mu$V/K). These findings underscore the significant potential of magnetic topological insulators for advanced thermoelectric applications.

Additionally, we show that predictions of thermopower based on Mott's relation become less reliable above the critical temperature, a consequence of the linear energy dependence of the topological surface states. Furthermore, the Wiedemann-Franz law is found to hold only at low temperatures; deviations at higher temperatures are attributed to strong spin-orbit interactions and the presence of magnetic clusters. This work underscores the promise of magnetic TIs for thermoelectric applications and paves the way for future research.

\section{Exchange interactions and magnetic clustering on the surface of magnetic TIs}\label{sec1}

\begin{figure}
	\centering
	\includegraphics[width=0.5\textwidth]{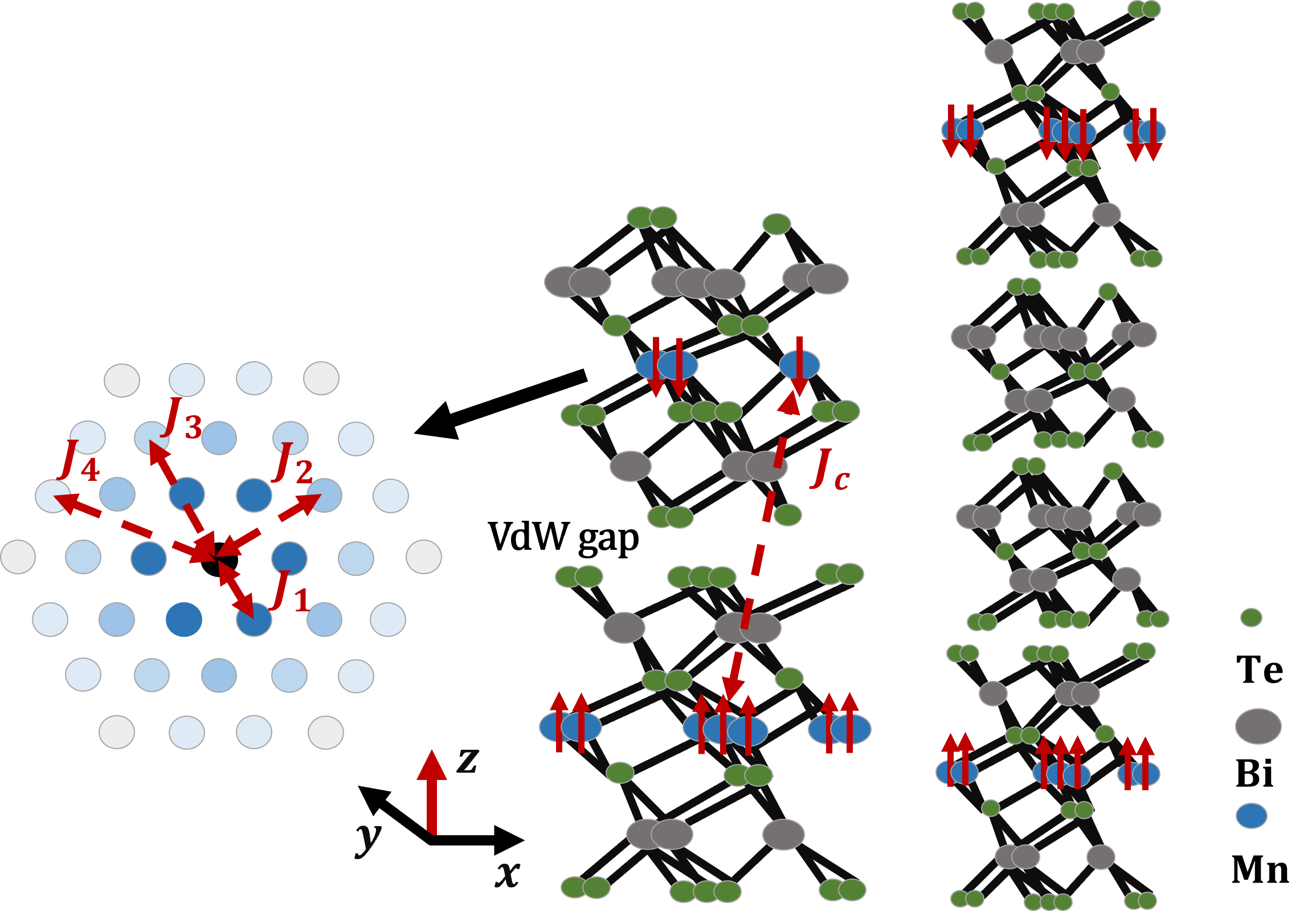}
	\caption{Left: A triangular lattice of ${\rm Mn}$ atoms, with exchange constants \(J_1\) to \(J_4\) representing interactions between Mn atoms with spin $ \tilde{s} = 5/2$ from NN to fourth NNs. In \(\rm MnBi_2Te_4\), they are characterized by the parameters $\tilde{s} J_1\approx 0.30$, $\tilde{s} J_2\approx -0.083$, $\tilde{s} J_3 \approx 0$, and $\tilde{s} J_4 \approx 0.023$ millielectron-Volt (${\rm meV}$), as reported in Ref. \cite{PRL_MagneticInteractions}. Middle: the lattice structure of the antiferromagnetic \(\rm MnBi_2Te_4\), which consists of two septuple layers separated by a van der Waals gap, with \(J_c\) representing the interlayer interaction between magnetic atoms. Right: the lattice structure of the ferromagnetic TI \(\rm MnBi_6Te_{10}\), in which two \(\rm Bi_2Te_3\) layers are placed between two \(\rm MnBi_2Te_4\) layers.}
	\label{fig: lattice}
\end{figure}

Before exploring the magnetic properties of magnetic TIs, we briefly review their typical lattice structure. Intrinsice magnetic TIs, wether antiferromagentic such as the chalcogenide semiconductors \(\rm MnBi_{2}Te_{4}\) and \(\rm MnBi_4Te_7\), or ferromagentic like \(\rm MnBi_6Te_{10}\) are composed of \(\rm Bi_2Te_3\) and \(\rm MnTe\) substances. The former features a layered structure formed by the stacking of septuple (Te-Bi-Te-Mn-Te-Bi-Te) triangular layers\cite{review_onMTI2021}, while the latter compounds consist of \(\rm MnBi_{2}Te_{4}\) units bridged by one or two \(\rm Bi_2Te_3\) layers (see Fig. \ref{fig: lattice}).

All atoms in the outermost septuple layer, which separates the TI from its environment, contribute to the surface properties. The magnetic atoms in this layer interact with surface Dirac electrons, thereby influencing the surface transport properties.

Inelastic neutron scattering experiments on \(\rm MnBi_{2}Te_{4}\) single crystals have demonstrated that the exchange interactions between Mn atoms are well-described by the Heisenberg spin Hamiltonian with uniaxial single-ion anisotropy \cite{PRL_MagneticInteractions, Li2021}. These measurements also reveal significant interlayer exchange interactions and substantial lifetime broadening. 

The exchange interactions between Mn atoms with spin $ \tilde{s} = 5/2$ are described by the following exchange Hamiltonian \cite{PRL_MagneticInteractions}:
\begin{equation}
	H=-\sum_{(i, j)_{\|}} J_{i j} \tilde{\mathbf{s}}_i \cdot \tilde{\mathbf{s}}_j - J_c \sum_{\langle i, j\rangle_{\perp}} \tilde{\mathbf{s}}_i \cdot \tilde{\mathbf{s}}_j - D \sum_i (\tilde{s}_i^z)^2.
	\label{eq:H1}
\end{equation}
This Hamiltonian is derived from fitting neutron scattering data to a local-moment Heisenberg model, limited to fourth-nearest neighbors within layers of Mn atoms. The first sum accounts for intra-triangular-layer interactions from nearest-neighbor (NN) to fourth NNs, the second sum covers interlayer NN interactions, and the final term represents the uniaxial single-ion anisotropy. The uniaxial anisotropy is in the $z$-direction, normal to the surface of the magnetic TI (see the figure \ref{fig: lattice}). This anisotropy significantly influences the transport properties of the TI, making the relaxation time isotropic and thereby eliminating anisotropic magnetoresistance.

Numerous experiments \cite{GMR1,GMR2,phase1,phase2}, particularly those investigating giant magnetoresistance (GMR) phenomena \cite{GMR2}, have demonstrated that spins orientation in magnetic materials significantly influences their electrical resistivity. In particualr, they reveal that the resistivity increases in the paramagnetic phase where the spins are randomly oriented. This behavior is attributed to the formation of magnetic clusters, whose size and number are temperature-dependent. In the following sections, we demonstrate that magnetic clustering on the surface of magnetic TIs, significantly influences the transport properties of the system. We define magnetic clusters within the Ising and Heisenberg models separately, and determine their size and number's temperature dependence using Monte Carlo (MC) simulations.

%%%%%%%%%%%%%%%%%%%%%%%%%%%%%%%%%%%%%%
\subsection{Magnetic clustering on the surface of TIs: Ising Hamiltonian}

Before delving into the original system, we model the interaction of magnetic atoms on the surface of magnetic TIs using a classical Ising Hamiltonian:  $H = -J_1\tilde{s}^2 \sum_{\langle i,j \rangle_{\|}} n_i n_j-J_c\tilde{s}^2 \sum_{\langle i,j \rangle_{\perp}} n_i n_j$, where \(J_1>0\) is NN ferromagnetic exchange constant, $J_c$ is NN inter-layer exchange coupling, and \(n_i\) denotes the classical spin at the lattice site $i$, taking values of \(\pm 1\). Given the Ising-like anisotropy in magnetic TIs, the insights from this model can be valuable for understanding the original system.

The two-dimensional Ising model on a triangular lattice with ferromagnetic exchange interaction exhibits two distinct phases \cite{Zhi-Huan2009}. At low temperatures, the system resides in a ferromagnetic phase characterized by non-zero magnetization. As the temperature increases, thermal fluctuations disrupt the long-range magnetic order. At a critical temperature, this order is destroyed, and the system undergoes a phase transition to a paramagnetic state. In the presence of inter-layer NN exchange coupling, whether ferromagnetic ($J_c>0$) or antiferromagnetic ($J_c<0$), the surface critical temperature is given by \(k_{\rm B}T_c \approx 4.2 J_1\tilde{s}^2\), where we choose $|J_c|/J_1=0.18$. 

In the ferromagnetic phase, magnetic clusters are formed by spins antialigned with the lattice magnetization. In the paramagnetic phase, although the magnetization is zero and long-range order is absent, domains with short-range order can exist at lower temperatures. In this phase, clustering occurs based on maximum exchange energy, with spins that have the highest energy grouping together into clusters.

\begin{figure}
	\centering
	\includegraphics[width=0.53\textwidth]{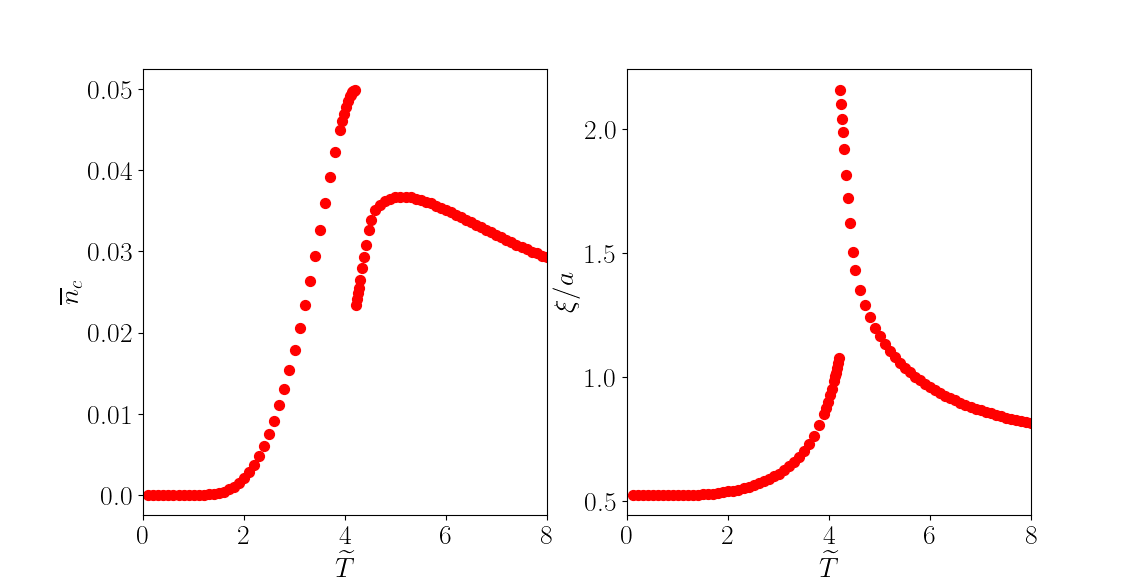}
	\caption{The MC results (Ising Hamiltonian) for the number of magnetic clusters per unit area, $\bar{n}_c$ (left), and the size of magnetic clusters, $\xi/a$ (right), as functions of temperature $\widetilde{T}=k_{\rm B}T/(J_1\tilde{s}^2)$. Here, $k_{\rm B}$ denotes the Boltzmann constant, $\tilde{s}$ represents the spin of the magnetic atoms, $J_1$ is the NN exchange interaction, and $a$ is the triangular lattice constant. The interaction of magnetic atoms on the surface of the magnetic TI is described by the Ising Hamiltonian. The number and size of clusters are independent of the nature of the interlayer interaction, whether antiferromagnetic ($J_c<0$) or ferromagnetic ($J_c>0$). In our analysis, the interlayer coupling $|J_c|$ is scaled to $J_1$ and set to 0.24, as referenced in \cite{PRL_MagneticInteractions}. The critical temperature is approximately $\widetilde{T}_c \simeq 4.2$ for both the antiferromagnetic and ferromagnetic cases.}
	\label{fig: xiI}
\end{figure}

Using MC simulations, we computed the size and number of clusters at various temperatures. To achieve equilibrium at a given temperature $T$, we used the final configuration from the last MC step at a lower temperature $T^\prime=T-\delta T$. To ensure thermal equilibrium, we performed $10^5$ MC steps per spin, followed by statistical averaging over an additional $10^5$ MC steps per spin. Our results for the number of clusters and their mean size are presented in Fig. \ref{fig: xiI}. 
As shown, for both ferromagnetic ($J_c>0$) and antiferromagnetic ($J_c<0$) interlayer exchange couplings, $\xi$ and $n_c$ are small at low temperatures, increase with rising temperature, peak at the critical temperature $T_c$, and then decrease with further increases in temperature. 

%%%%%%%%%%%%%%%%%%%%%%%%%%%%%%%%%%%%%%%%
\subsection{Magnetic clustering on the surface of TIs: Heisenberg Hamiltonian}

In this section, we consider the interaction of magnetic atoms on the surface of the TI to be described by the Heisenberg exchange model in Eq. (\ref{eq:H1}). It has been demonstrated that this model on a triangular spin lattice exhibits various long-range orders. In a triangular lattice with ferromagnetic exchange interactions between NN spins ($J_1>0$) and antiferromagnetic exchange interactions between next-NN spins ($J_2<0$), the ground state is ferromagnetic when $J_2 \ll J_1$. However, it transitions to an incommensurate short-range ordered phase at $|J_2/J_1|=1/3$ \cite{Matsubara1996}.

In most magnetic materials with magnetic ions arranged in a triangular lattice, such as $\rm LiNiO_2$ \cite{Matsubara1996}, the ratio of coupling constants $J_2/J_1<1/3$ results in a ferromagnetic ground state. However, at non-zero temperatures, thermal fluctuations disrupt this magnetic order, leading to a paramagnetic phase at all temperatures, as predicted by the Mermin-Wagner theorem.

In intrinsic antiferromagnetic TIs like $\rm MnBi_2Te_4$ and $\rm MnBi_4Te_7$, uniaxial single-ion anisotropy breaks the SU(2) symmetry, thereby allowing the surface of the TI to maintain ferromagnetic order even at finite temperatures. However, as the temperature increases, thermal fluctuations become more significant, eventually causing a transition to a paramagnetic phase at a critical temperature $T_c$. Additional ferromagnetic interactions further stabilize the ferromagnetic order within the triangular layer.

When the interactions among magnetic atoms on the surface of the TI are modeled using the Heisenberg model (Eq. (\ref{eq:H1})), the lattice spins can assume a variety of orientations. At zero temperature, due to the presence of easy-axis anisotropy, the spins are fully aligned along the $z$-axis (the easy axis), resulting in a magnetization per spin of $m^z = \tilde{s}$, where $\tilde{s}$ denotes the spin magnitude of the magnetic atoms. As temperature increases, some spins, referred to as defect spins, become oriented opposite to the overall magnetization, leading to the formation of magnetic clusters.

A spin can be a member of a cluster if it meets the following conditions: (1) its $z$-component is opposite to the lattice magnetization, and (2) it is a NN to a spin in the cluster. Once these criteria are met, the spin becomes a cluster member with the Wolff probability $P=1-\exp(-2E/k_{\rm B}T)$ \cite{wolff1,Coniglio}, where $E$ refers to the energy of links connecting the spins to their first to fourth NNs.

To determine the size and number of clusters in the Heisenberg model, we performed the following steps:
\begin{enumerate}
	\item{At a specified temperature $T$, allow the system to reach equilibrium.}
	\item{Randomly select a spin and identify the cluster it belongs to.}
	\item{Add neighboring spins to the cluster using the Wolff probability. Below the critical temperature, because the surface magnetization is nonzero, a spin can belong to a cluster with Wolf probability, provided it is not aligned with the overall magnetization.}
	\item{Replace the spins in the cluster with their average spin.}
	\item{Repeat steps 3 and 4 for the newly added spins.}
	\item{Repeat steps 2 to 5 until all spins have been visited.}
	\item{Record the average number and size of clusters \cite{hoshen}, along with the direction indicated by the angles $\theta_s$ and $\phi_s$, for analytical calculations.}
	\item{Increase the temperature by $\delta T$ and use the final state from the previous temperature as the initial state.}
	\item{Repeat steps 1 to 8 for the desired temperature range.}
\end{enumerate}
\begin{figure}
	\centering
	\includegraphics[width=0.45\textwidth]{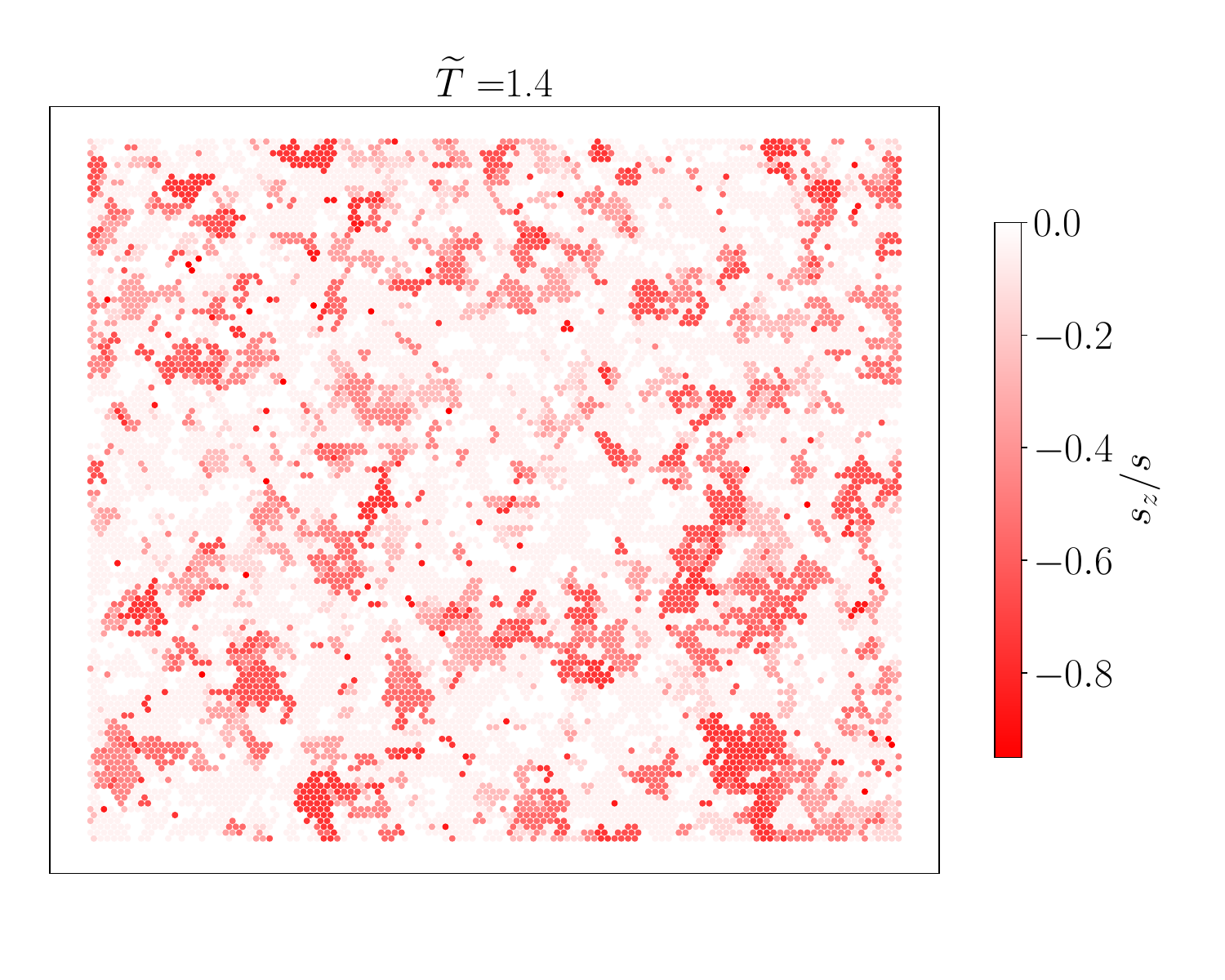}
	\includegraphics[width=0.45\textwidth]{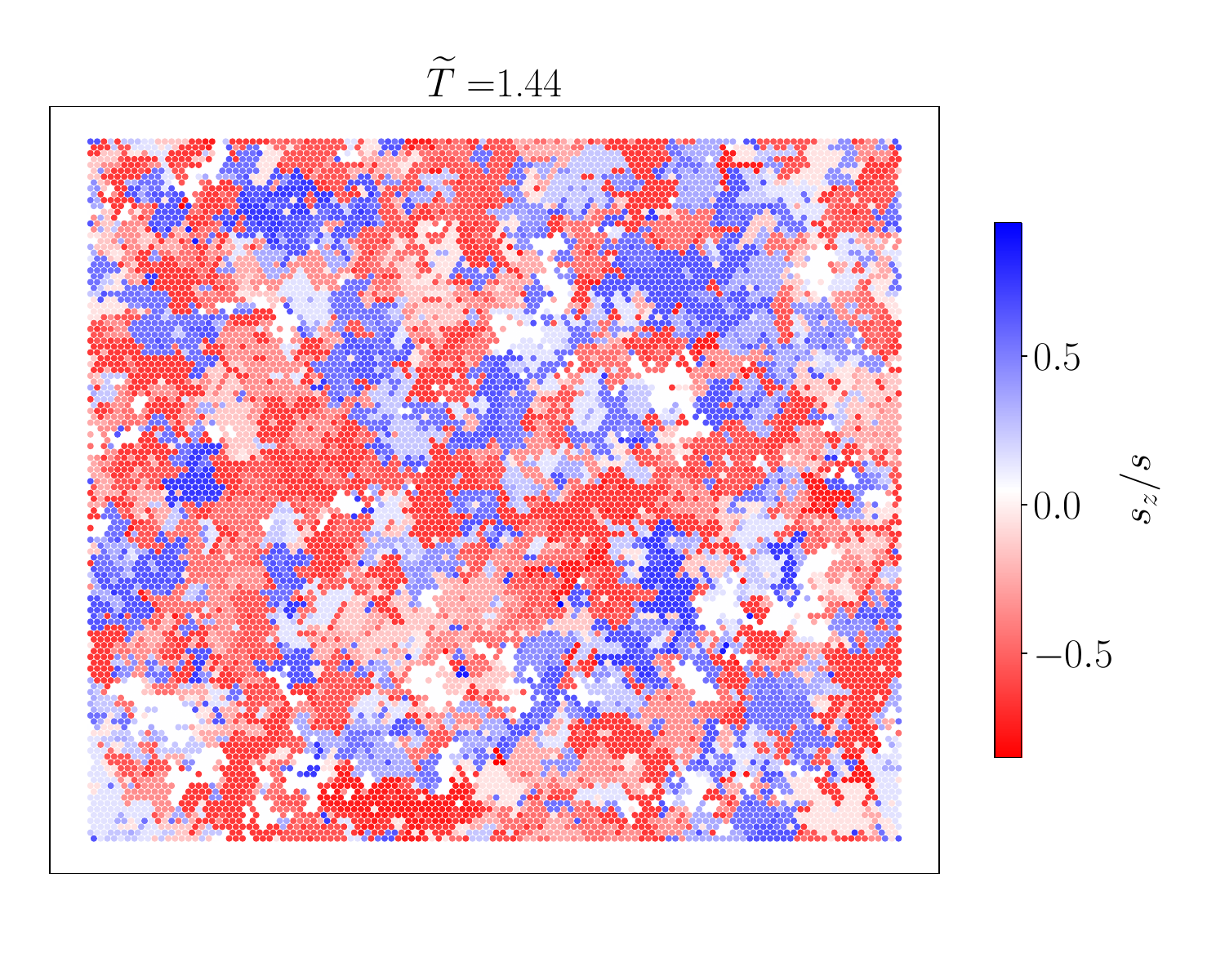}
	\includegraphics[width=0.45\textwidth]{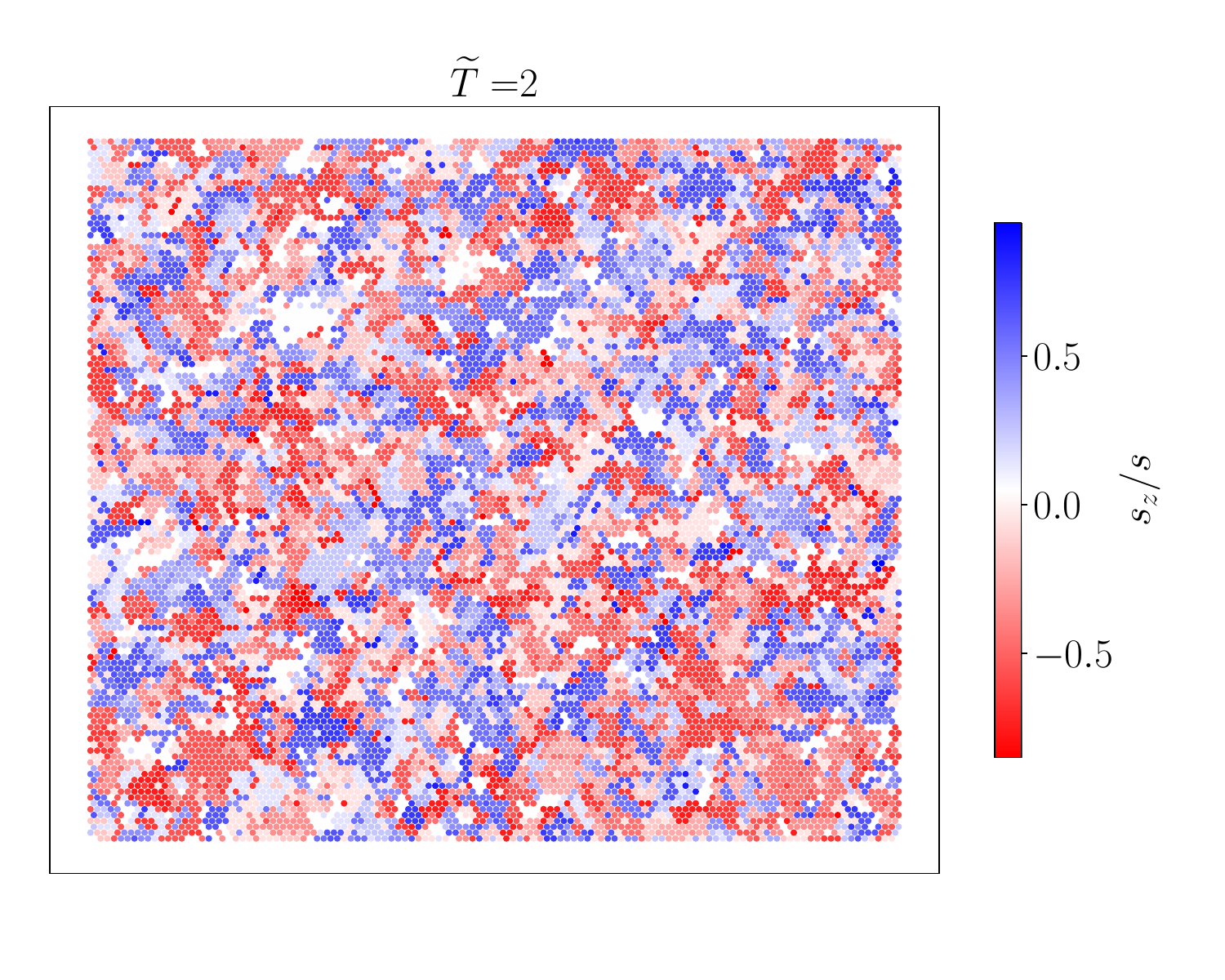}
	\caption{Clustering of magnetic atoms located on the triangular lattice sites on the TI surface, driven by the Heisenberg interaction (see Eq. (\ref{eq:H1})). The different clusters are distinguished by the ratio $s_z/s$. Here, $\widetilde{T}=k_{\rm B}T/(J_1\tilde{s}^2)$, $J_1$ is the NN exchange coupling, and the exchange couplings $J_2$ and $J_4$, the uniaxial single-ion anisotropy $D$, and the interlayer exchange $J_c$ are scaled to $J_1$ with values set to $-0.28$, $0.08$, $0.4$, and $-0.18$, respectively. Different colors denote various cluster types. In the ordered phase, below the critical temperature $\widetilde{T}_c=1.42$, all clusters exhibit a negative spin value $s_z<0$ (surface magnetization is along the $+z$-direction). As the system crosses the critical point, clusters with $s_z>0$ emerge on the surface. Above the critical point, increasing temperature results in smaller cluster sizes.}
	\label{fig: xiH}
\end{figure}

Unlike the Ising model, where the clustering mechanism differs in the ferromagnetic and paramagnetic phases, the Heisenberg model applies the same steps in the paramagnetic phase. In this phase, short-range ordered regions lead to a larger scattering amplitude compared to the ferromagnetic phase. Since neighboring spins with the same orientation exhibit a higher Wolff probability, this probability is also used to construct magnetic clusters in the paramagnetic regime.

Our MC simulations demonstrate that clusters display varying spin components along the $z$-axis, denoted as $s_z$ (see Fig. \ref{fig: xiH}). Accordingly, we classify these clusters based on their $s_z$ values or, equivalently, their polar angles $\theta_s$ relative to the $z$-axis. Each class is treated as an independent defect type, assumed to scatter electrons independently.

\begin{figure}
	\centering
	\includegraphics[width=0.4\textwidth]{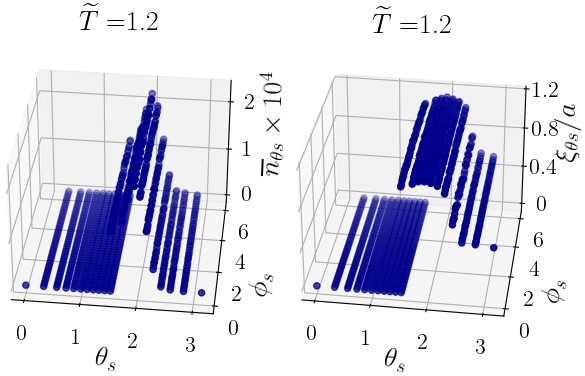}
	\includegraphics[width=0.45\textwidth]{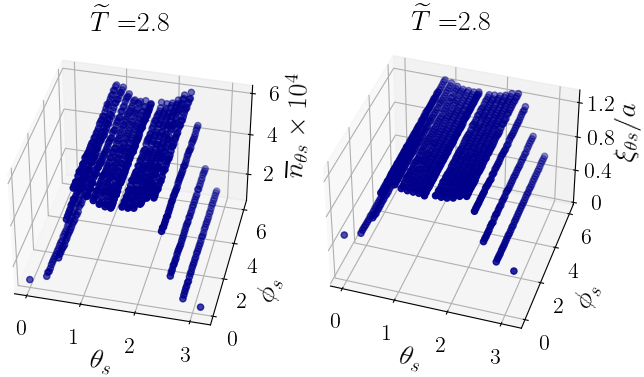}
	\includegraphics[width=0.5\textwidth]{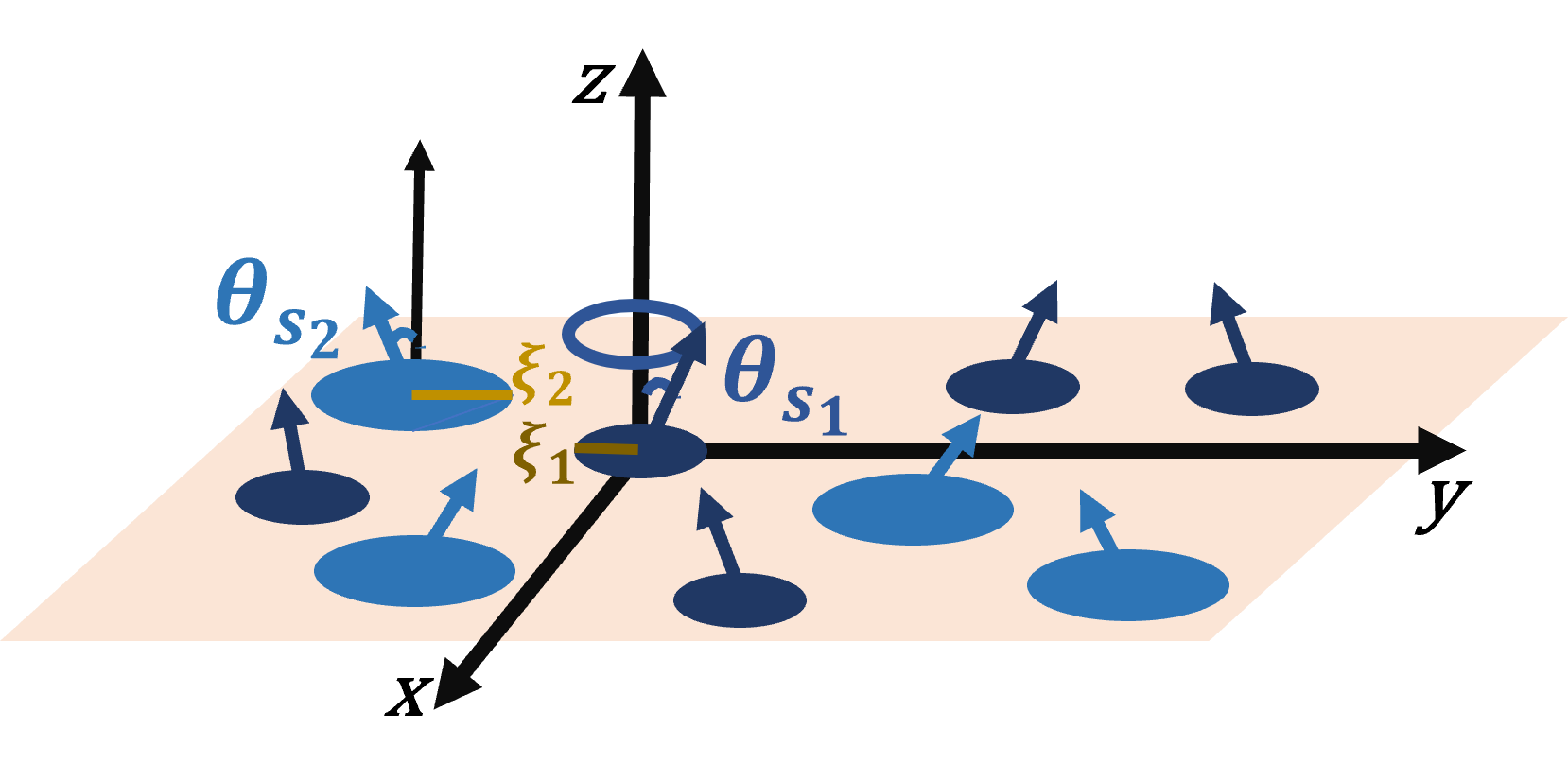}
	\caption{The MC results for the mean size and number of clusters of different types at temperatures, below ($\widetilde{T} = 1.2$) and above ($\widetilde{T} = 2.8$) the critical temperature ($\widetilde{T}_c = 1.42$). Here, $\widetilde{T} = k_{\rm B} T / (J_1 \tilde{s}^2)$, with $J_1$ being the NN exchange constant. When the interactions between magentic atoms are described by the Heisenberg Hamiltonian, various types of clusters form, characterized by their $s_z$ component or, equivalently, their polar angles $\theta_s$ relative to the $z$-axis (see also the MC results presented in Fig. \ref{fig: xiH}).
For each $\theta_s$, both the size and the number of clusters are independent of the azimuthal angle $\phi_s$ (above four plots). In the bottom panel, we schematically depict two types of clusters, categorized by their sizes and spin angles: $\xi_1$ with angle $\theta_{s_1}$ (dark blue), and $\xi_2$ with angle $\theta_{s_2}$ (light blue). The spins within clusters of the same type are oriented such that their in-plane components cancel each other out on average.}
	\label{fig: clustering-schematic}
\end{figure}

The simulations further demonstrate that the interplay between the isotropic Heisenberg interaction and the on-site anisotropy term in Eq. (\ref{eq:H1}) gives rise to symmetric spin configurations. In these configurations, clusters tend to orient such that their in-plane spin components cancel out on average, as depicted in Figure \ref{fig: clustering-schematic}. Specifically, for a cluster oriented at angles $(\theta_s, \phi_s)$, there exists a corresponding cluster at $(\theta_s, -\phi_s)$, resulting in a net spin component aligned along the $z$-axis. This component is quantified by $s_{z} = s \cos \theta_s$, where $s$ denotes the magnitude of the cluster's spin.

A schematic representation of this symmetry is shown at the bottom of Fig. \ref{fig: clustering-schematic}. Clusters sharing the same $\theta_s$ are color-coded uniformly, emphasizing their similar contributions to the scattering process.

\begin{figure}
	\centering
	\includegraphics[width=0.52\textwidth]{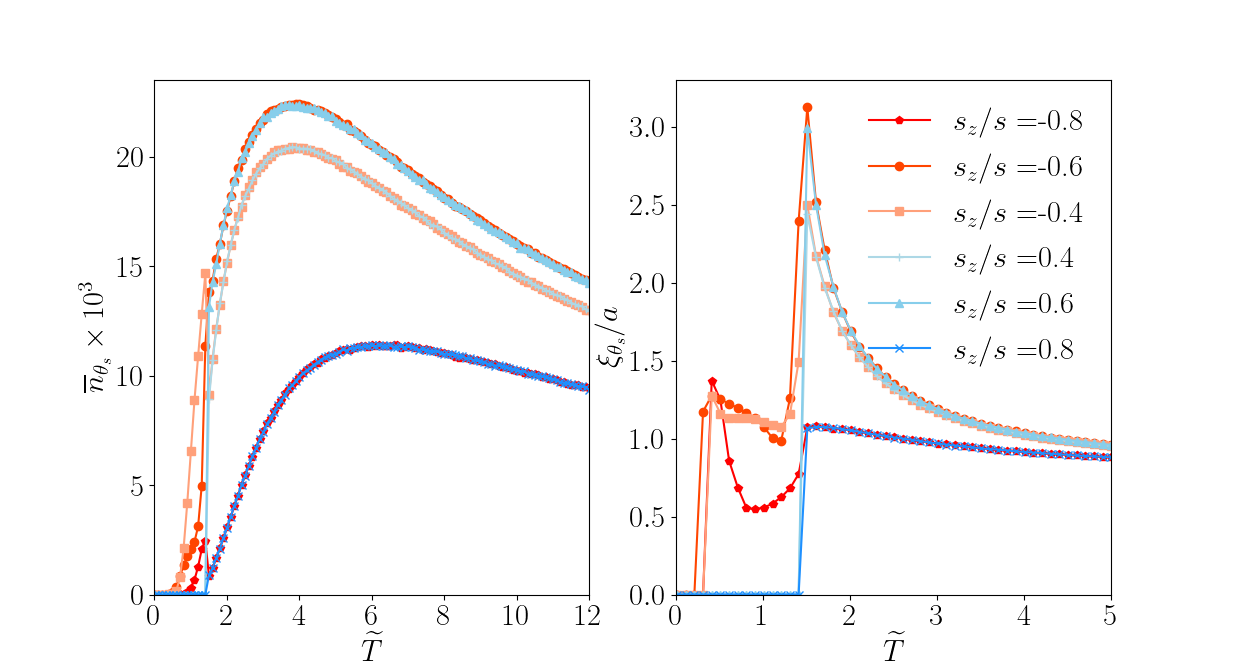}
	\caption{The MC results for the mean size and number of clusters of different types, labeled by $s_z/s=\cos\theta_s$, as functions of $\widetilde{T} = k_{\rm B} T / (J_1 \tilde{s}^2)$. The interaction of magnetic atoms on the surface of the magnetic TI is described by the Heisenberg Hamiltonian. The interlayer interactions are antiferromagnetic ($J_c<0$), with $|J_c|/J_1=0.18$. At temperatures below the critical temperature, only clusters with negative $s_z$, depicted in red, appear on the surface of the TI, while above the critical temperature, all types of clusters emerge.
	}
	\label{fig: Xi-n-HAnti}
\end{figure}
We have also plotted in Fig. \ref{fig: Xi-n-HAnti} the size and number of clusters of different types as a function of temperature. At temperatures below the critical temperature, only clusters with negative $s_z$, depicted in red, appear on the surface of the TI, while above the critical temperature, all types of clusters emerge. 

 In the Heisenberg model, the number of clusters as a function of temperature exhibits no discontinuity at the critical point, unlike in the Ising model. This difference stems from the fact that, in the Ising model, the criteria for defining clusters differ below and above the critical temperature. Specifically, below the critical point, clusters are determined based on magnetization, while above it, they are defined using link energy. In contrast, the Heisenberg model employs the Wolf probability for cluster definitions on both sides of the critical point, leading to a continuous behavior of the cluster number across the transition.

In the antiferromagnetic phase, magnetic clusters exhibit comparable numbers and sizes for spins with equal magnitudes of $|s_z|$. In fact, there is no notable difference between the magnetic clusters characterized by $|s_z|$ and those characterized by $-|s_z|$.

We performed our MC simulations initially for a finite number of stacked layers, each containing magnetic ions on a triangular lattice in the xy-plane. We considered a two-dimensional lattice with $120\times 120$ sites under periodic boundary conditions. To enhance computational efficiency and improve system simulation, we initially focused on a limited number of layers.

\begin{figure}
	\centering
	\includegraphics[width=0.5\textwidth]{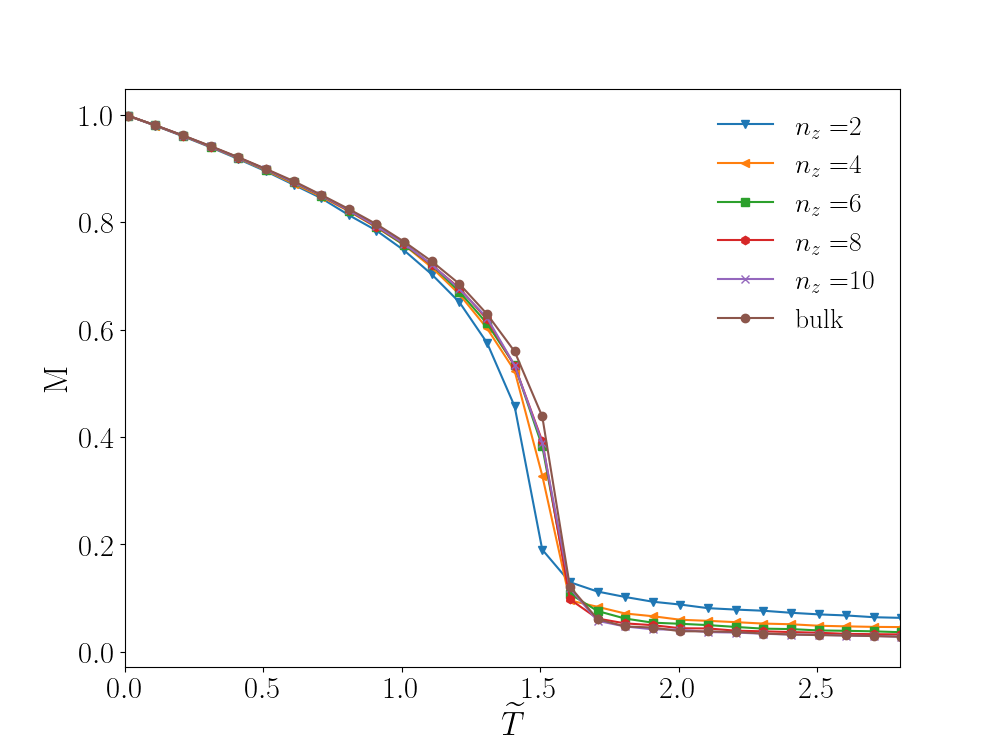}
	\caption{The bulk magnetization per spin as a function of $\widetilde{T} = k_{\rm B} T / (J_1 \tilde{s}^2)$ for various numbers of layers ($n_z$). Here, the intralayer interaction is given by the Heisenberg model with NN exchange interaction $J_1$, and the interlayer interaction is set ot $J_c/J_1=-0.18$. This result shows that a TI with $n_z=10$ magnetic layers can fully reflect the magnetic properties of the 3D TI.}
	\label{fig: Order}
\end{figure}

To investigate the impact of the number of layers, we systematically increased the number of layers and compared the results to a system with periodic boundary conditions along the $z$-axis. Our simulations show that when the number of layers is about ten, the magnetization behavior converges to that of a periodic system in the $z$-direction (see Fig. \ref{fig: Order}). 
This indicates that a ten-layer magnetic TI exhibits behaviors akin to those of 3D magnetic TIs, supporting previous research \cite{Otrokov2019,Zhao2021}. 

%%%%%%%%%%%%%%%%%%%%%%%%%%%%%%%%%%%
%%%%%%%%%%%%%%%%%%%%%%%%%%%%%%%%%%%

\section{Electron-Clusters scattering potential and relaxation time}\label{scattering}

The dynamics of Dirac electrons on the surface of magnetic TIs is described by the Bloch Hamiltonian:
\begin{equation}
	H_k=\hbar v_{\rm F}(\vec{k}\times\vec{\sigma})\cdot\hat{z}+ J_bM\sigma_z.
	\label{Eq: HDirac}
\end{equation}
Here, the unit vector $\hat{z}$ is taken to be perpendicular to the surface of the TI. The Fermi velocity is denoted by $v_{\rm F}$, and the momentum of Dirac electrons is expressed as $\vec{k} = k_{x}\hat{x} + k_{y}\hat{y}$. The Pauli matrix vector $\vec{\sigma}$ characterizes the spin degrees of freedom of the Dirac electrons. The term $J_bM$ represents the molecular field acting on the surface electrons, induced by magnetic atoms embedded in the bulk. This field opens an energy gap of magnitude $2J_bM$ in the surface state spectrum.
%This energy gap leads to phenomena such as skew scattering and the anomalous Hall effect. 

The bulk magnetization $M$ is temperature-dependent. At temperatures higher than the critical temperature of the bulk $T>T_c^{\text{bulk}}$, the magnetization is zero, resulting in gap-closing in the system. Therefore, when examining the thermoelectric properties of the magnetic TI, this temperature dependence should be considered in our formalism. The energy spectra and wavefunctions of the Dirac electrons are given by:
\begin{equation}
	\E_k=\pm\sqrt{(\hbar v_{\rm F}k)^2 + J_b^2M^2},
	\label{Eq: Ek}
\end{equation}
and
$\psi_{k}(\vec{r}) =\frac{1}{\sqrt{A}} e^{i\vec{k}\cdot\vec{r}}|u_{k\pm}\rangle$, with:
\begin{equation}\label{me_3}
	|u_{k+}\rangle = \begin{pmatrix}
		-i e^{-i\phi_k}\cos\frac{\eta_k}{2} \\
		\sin\frac{\eta_k}{2}
	\end{pmatrix},
	|u_{k-}\rangle = \begin{pmatrix}
		i e^{-i\phi_k}\sin\frac{\eta_k}{2} \\
		\cos\frac{\eta_k}{2}
	\end{pmatrix},
\end{equation}
where $k=|\vec{k}|=(k_x^2+k_y^2)^{1/2}$, $\cos\eta_k=J_bM/|\E_k|$ and $\phi_k=\tan^{-1}(k_y/k_x)$.

As discussed in the previous section, exchange interactions between magnetic atoms lead to magnetic clustering on the surface of the TI. These clusters are randomly distributed, uncorrelated, and scatter electrons, resulting in surface magnetoresistance. To determine the scattering rate, we express the scattering potential in terms of the clusters' mean size $\xi$ as follows \cite{Zarezad2018, diep2011}:
\begin{equation}
	\label{eq: e-c-interaction}
	H_{\sigma s} = J_0\exp\left(-|\vec{r}-\vec{R}|/\xi\right)\vec{\sigma}_{\vec{r}}\cdot\vec{s}_{\vec{R}},
\end{equation}
where $\vec{s}$ is the spin of the cluster, and $J_0$ is the exchange coupling between the Dirac electron at $\vec{r}$ and the cluster at $\vec{R}$. 

Expressing the scattering potential in terms of the clusters' mean size allows us to examine the effects of temperature and phase transitions on the surface's transport properties.

In electron scattering from a magnetic impurity, both the spin and size of the magnetic impurity are typically considered in the scattering potential. This includes an $s-d$ term and a Dirac delta term for charge scattering. However, in the scattering potential described in Eq. (\ref{eq: e-c-interaction}), both terms are already incorporated.

When the interaction between magnetic atoms on the surface of the TI is described by the Ising Hamiltonian, the spins $\vec{s}$ of the clusters are perpendicular to the TI's surface. In this scenario, the scattering rate is isotropic and depends solely on $\Delta\vec{k}=\vec{k}-\vec{k'}$. The relaxation time can be derived from:
\begin{equation}
	\label{eq: tauIsotropic}
	\frac{1}{\tau_{k}} = A\int\frac{d^{2}k'}{4\pi^{2}}w(\vec{k},\vec{k'})(1-\cos\Delta \phi),
\end{equation}
where $w(\vec{k},\vec{k'})$ is the transition rate between the two eigenstates $|\vec{k}\rangle$ and $|\vec{k'}\rangle$ of the Hamiltonian (\ref{Eq: HDirac}), and $\Delta\phi=\phi_k-\phi_{k'}$. Using Fermi golden rule the transition rate simply reads:
\begin{equation}
	\label{eq: rate}
	w(\vec{k},\vec{k'}) = \frac{2\pi}{\hbar}n_c|T_{\vec{k},\vec{k'}}|^{2}\delta(\E _{k}-\E _{k'}),
\end{equation}
where $n_{c}$ is clusters number. In various systems, \(J_0\) typically amounts to one or two orders of magnitude less than the Fermi energy. For magnetic TIs, \(J_0\) is on the order of 10 ${\rm meV}$ (for example, in \(\mathrm{MnBi_2Te_4}\), \(J_0 \sim 25\) meV\cite{J0_2022}), reflecting a weak scattering potential that justifies the use of the first Born approximation.
Within this approximation, the $T$-matrix is obtained as follows:
\begin{equation}
	T_{\vec{k},\vec{k'}}=J_{0}s\sum_{\vec{r},\vec{R}}\langle\vec{k}|\exp\left(-|\vec{r}-\vec{R}|/\xi\right)\int d^{2}r |\vec{r}\rangle \langle\vec{r}|\vec{k'}\rangle.
\label{eq:T1}
\end{equation}
The relaxation time is therefore obtained as follows (for detailed calculations, refer to the appendix):
\begin{equation}
	\label{eq: tau1}
	\frac{1}{\tau_k}=\frac{1}{\tau_0} \bar{n}_c k\xi (\xi/a)^3 \frac{3+(1+4k^{2}\xi^{2})\cos^{2}\eta_k}{\sin\eta_k (1+4k^{2}\xi^{2})^{5/2}},
\end{equation}
where $\tau_0= \frac{\sqrt{3}\hbar^2v_{\rm F}}{2\pi^2 J_0^2s^2a}$ is a constant with the dimension of time, and $\bar{n}_c=n_c/N$ represents the average number of clusters per triangular lattice site ($N$). The constant $\sqrt{3}a^2/2$ is the area occupied by each magnetic atoms positioned on a triangular lattice. In the case of square lattice this contant reduces to $a^2$.

The situation differs when the interaction of magnetic atoms on the surface of the TI is described by the Heisenberg Hamiltonian in Eq. (\ref{eq:H1}). In this case, as discussed, various types of clusters form on the TI surface, each identified by its spin angle relative to the $z$-axis ($\theta_s$). These clusters act as different types of defects, scattering electrons independently. The relaxation time of Dirac electrons scattering off clusters with spin angle $\theta_s$ is given by (for detailed calculations, refer to the appendix):
\begin{equation}
	\label{eq: tau2}
	\frac{1}{\tau_{k,\theta_s}} =\frac{1}{\tau_0}\bar{n}_{\theta_s} k \xi_{\theta_s} \left(\frac{\xi_{\theta_s}}{a}\right)^3\frac{g(\theta_s,\eta_k)}{\sin\eta_k (1+4k^{2}\xi_{\theta_s}^{2})^{5/2}},
\end{equation}
where 
\begin{align}
g(\theta_s,\eta_k)&=\cos^2\theta_s[3+(1+4k^2\xi_{\theta_s}^2)\cos^2\eta_k]\nonumber\\
&+2(1+k^2\xi_{\theta_s}^2)\sin^2\eta_k\sin^2\theta_s.
\end{align}
Here, $\xi_{\theta_s}$ is the mean size of clusters with spin angle $\theta_s$, and $\bar{n}_{\theta_s}$ is their number per unit area.

The overall relaxation time for Dirac electrons, considering all cluster types, is obtained via Matthiessen's rule:
\begin{equation}
	\label{Eq: Matth}
	\frac{1}{\tau_k}=\sum_{\theta_s}\frac{1}{\tau_{k,\theta_s}}.
\end{equation}

The relaxation times in both the Ising and Heisenberg models depend on temperature through the temperature dependence of the cluster size and number, as well as the surface gap ($2J_bM$). In layered compounds such as $\rm MnBi_2Te_4$ and $\rm MnBi_4Te_7$, the interlayer interactions are antiferromagnetic\cite{Shao2021}, and the bulk has A-type Neel order resulting in vanishing bulk magnetization and gapless surface states. Conversely, in compounds like $\rm MnBi_6Te_{10}$, ferromagnetic interlayer interactions magnetize the bulk\cite{Shao2021}, leading to a gap in the surface states. Unlike the cluster size and number, which increase with rising temperature, the surface gap decreases due to the reduction in bulk magnetization. At the bulk critical point the magnetization becomes zero, causing the surface gap to close ($\cos\eta_k=0$). For temperatures above $T_c^{\text{bulk}}$, the gap remains closed, and the temperature dependence of the relaxation times is primarily influenced by the behavior of the cluster size and number.
\begin{figure}
	\centering
	\includegraphics[width=0.45\textwidth]{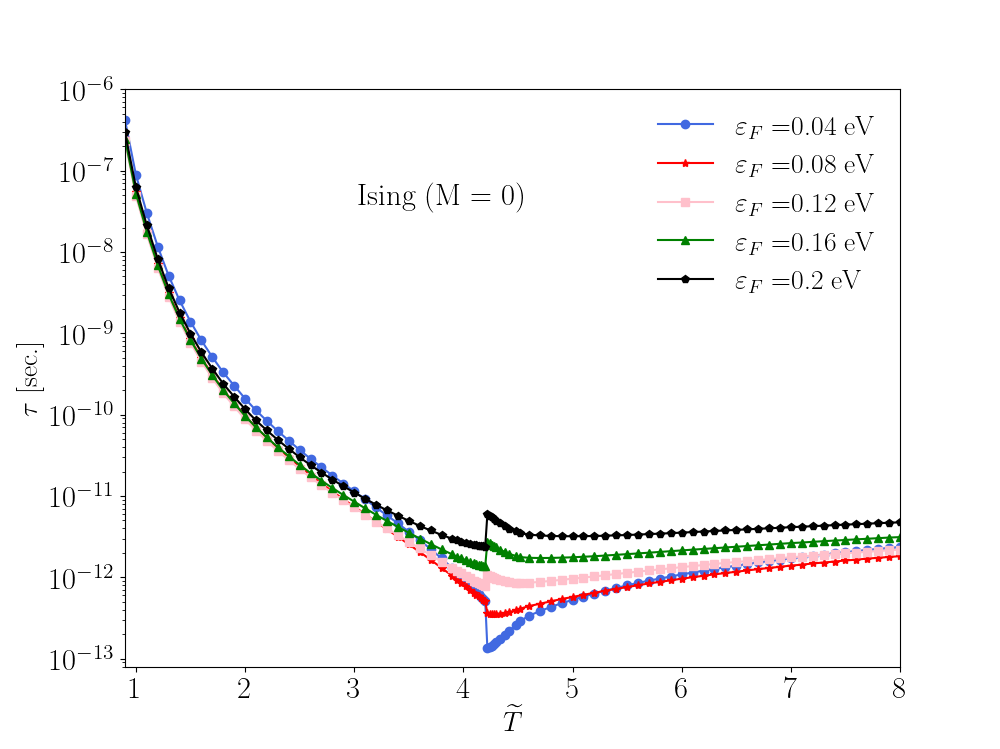}
	\includegraphics[width=0.45\textwidth]{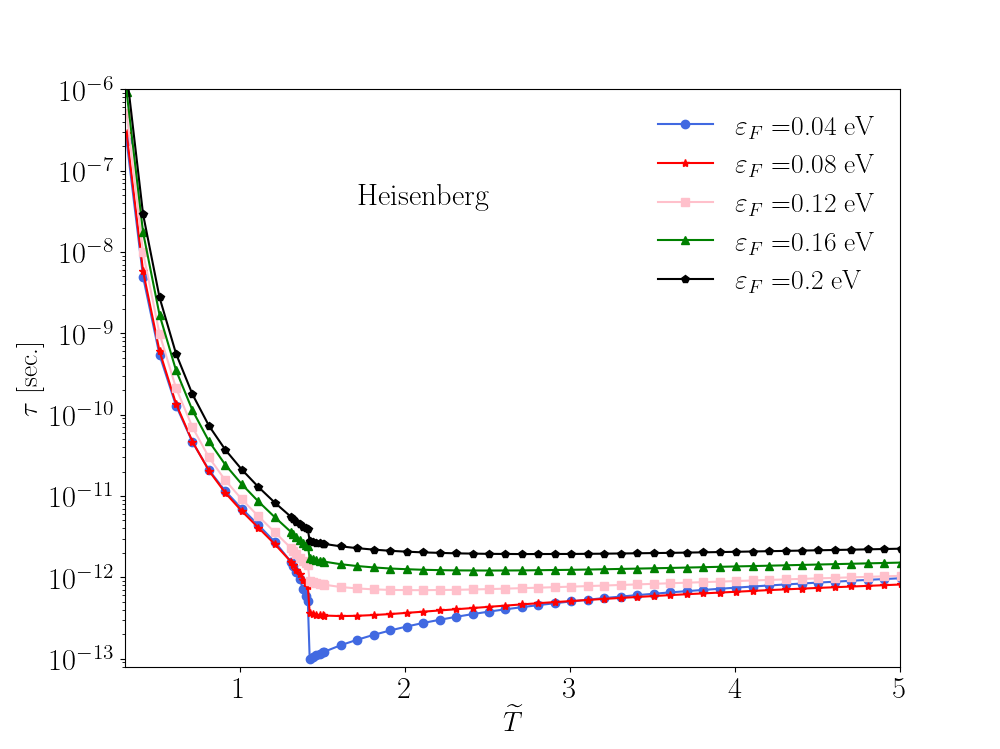}
	\caption{The relaxation time of Dirac electrons as a function of $\widetilde{T}=k_{\rm B}T/(J_1\tilde{s}^2)$, for various Fermi energies. The exchange interaction between magnetic atoms are modeled using the Ising (up) and Heisenberg (down) Hamiltonians. In the Heisenberg case, $J_1$ represents the NN exchange coupling, while $J_2$, $J_4$, the uniaxial single ion anisotropy $D$, and the interlayer exchange $J_c$ are scaled to $J_1$, with values set to $-0.28$, $0.08$, $0.4$, and $-0.18$, respectively. Here, we set $\tilde{s} J_1 = 0.3$ meV, consistent with parameters for manganese bismuth telluride.}
	\label{fig: relaxationT}
\end{figure}

Figure \ref{fig: relaxationT} illustrates the relaxation time of gapless Dirac electrons ($M=0$) as a function of temperature, for the parameters $J_0=25~\text{meV}$\cite{J0_2022}, $s=5/2$, $v_{\rm F}=5.5\times 10^{5}\text{m/s}$, $a=4.33 \text{{\AA}}$, and $\tau_0\simeq1.3\times 10^{-14}~\text{s}$. 
The dependence of relaxation times on temperature exhibits similar qualitative behavior in both the Ising and Heisenberg models for temperatures below the critical point. At low temperatures, the magnetic clusters are small in size and number, resulting in low scattering amplitudes and consequently long relaxation times. As temperature increases toward the surface critical temperature, both the clusters size and number grow, leading to a rapid decrease in relaxation time. At the surface critical temperature, the cluster size and number reach their maximum values, corresponding to a minimum in relaxation time.

At the critical temperature, the models diverge in their relaxation time behaviors. In the Heisenberg model, relaxation times exhibit an abrupt decrease across all Fermi energies, indicating a rapid dynamical response. Conversely, in the Ising model, the relaxation time behavior is non-uniform: it displays a sudden increase at high Fermi energies and a sudden decrease at low Fermi energies. This contrasting behavior can be attributed to the dominant factors influencing relaxation dynamics. At low Fermi energies, the relaxation time is primarily governed by the size of magnetic clusters, whereas at high Fermi energies, the number of clusters exerts a more significant influence. These differences highlight the distinct nature of magnetic correlations and thermal fluctuations in the two models near criticality.

%%%%%%%%%%%%%%%%%%%%%%%%%%%%
%%%%%%%%%%%%%%%%%%%%%%%%%%%%

\section{Thermoelectric properties}\label{transport}

At a given temperature $T$, the efficiency of thermoelectric materials is characterized by the figure of merit $zT=\frac{S^2\sigma}{\kappa}T$, where $\sigma$, $S$, and $\kappa$ represent electrical conductivity, thermoelectric power factor (thermopower), and thermal conductivity, respectively. These properties vary with temperature due to the temperature dependence of the relaxation time $\tau_k$. They are expressed in terms of $\tau_k$ as:
\begin{equation}
	\gamma^{(i)}(T) = \int\frac{e^{2-i}}{4 \pi} \tau_k(\E _k-\mu)^{i} v_k^2\left(\frac{-\partial f^0}{\partial \E _k}\right) k d k,
	\label{eq: gamma}
\end{equation}
where $e=-1.6\times 10^{-19}~\text{C}$ is the electron charge, $v_k$ is band velocity, $\mu$ is the chemical potential, and $f^0$ is Fermi-Dirac distribution function. The functions $\gamma^{(i)}$ with $i=0, 1$, and 2 correspond to $\gamma^{(0)}=\sigma$, $\gamma^{(1)}=\sigma S T$, and $\gamma^{(2)}=\kappa T$. 
The temperature dependence of the chemical potential $\mu$ is as:
\begin{equation}
	\mu(T)=\E _{\rm F} \left[1-\frac{\pi^2}{3}\left(\frac{k_{\rm B}T}{\E _{\rm F}}\right)^{2}\right]^{1/2},
	\label{eq: mu}
\end{equation}
where $\E_{\rm F}$ is the Fermi energy. 
As the surface density of states is linear with surface energy, i.e. $D(\E )=\E /(2\pi\hbar^2v_{\rm F}^2)$, the above expression for the chemical potentail is exact. 

At sufficiently low temperatures, the chemical potential approximates the Fermi energy, $\mu \simeq \E _{\rm F}$, allowing the derivative of the Fermi-Dirac distribution function to be approximated by the Dirac delta function, $\delta(\E _k - \E _{\rm F})$. In this regime, the electrical conductivity is given by $\sigma=c~\tau_{k_{\rm F}}$, where $c$ is a constant, while both the thermal conductivity and the thermopower vanish. As temperature increases, the chemical potential decreases according to Eq. (\ref{eq: mu}), reaching zero at $k_{\rm B}T=\sqrt{3}\E _{\rm F}/\pi \simeq \E _{\rm F}/2$. For Fermi energies ranging from 10 to $200~\text{meV}$, the corresponding temperatures range from $100~\text{K}$ to $2000~\text{K}$, some of which are significantly higher than the stability temperature of magnetic TIs. The van der Waals interactions in magnetic TIs ensure the stability of their crystal structure up to approximately $500-600~\text{K}$; however, beyond this temperature, these interactions diminish in significance, leading to a transition of the crystal into flakes \cite{flakes1,flakes2}.

Various proposals have been made to maintain material stability at high temperatures, one of which involves applying external pressure \cite{Htemperature1,Htemperature2}. It has been demonstrated that applying pressure can raise the stability temperature to $900~\text{K}$. Therefore, our subsequent investigations into the effects of high temperatures are justified.

%%%%%%%%%%%%%%%%%%%%%%%%%%%%%%%%%%
\subsection{Ising case: Antiferromagnetic interlayer interactions}

When the interaction between magnetic atoms on the surface of magnetic TIs is described by the Ising model, the relaxation time of Dirac electrons is given by Eq. (\ref{eq: tau1}). For intrinsic antiferromagnetic TIs like \(\rm MnBi_2Te_4\), the bulk exhibits Neel order with a finite staggered magnetization below the Neel temperature $T_{\rm N}=25~\text{K}$, transitioning to a paramagnetic phase above $T_{\rm N}$. In both phases, the magnetization $M$ is zero, resulting in a vanishing surface gap. Consequently, in this scenario, $\cos\eta_k=0$, and the relaxation time simplifies to:
\begin{equation}
	\frac{1}{\tau_k}=\frac{1}{\tau_0} \bar{n}_c k\xi (\xi/a)^3 \frac{3}{(1+4k^2\xi^2)^{5/2}}.
\end{equation}
As a result, the electrical and thermal resistivities can be determined by numerically solving the integrals presented in Eq. (\ref{eq: gamma}).
\begin{figure}
	\centering
	\includegraphics[width=0.5\textwidth]{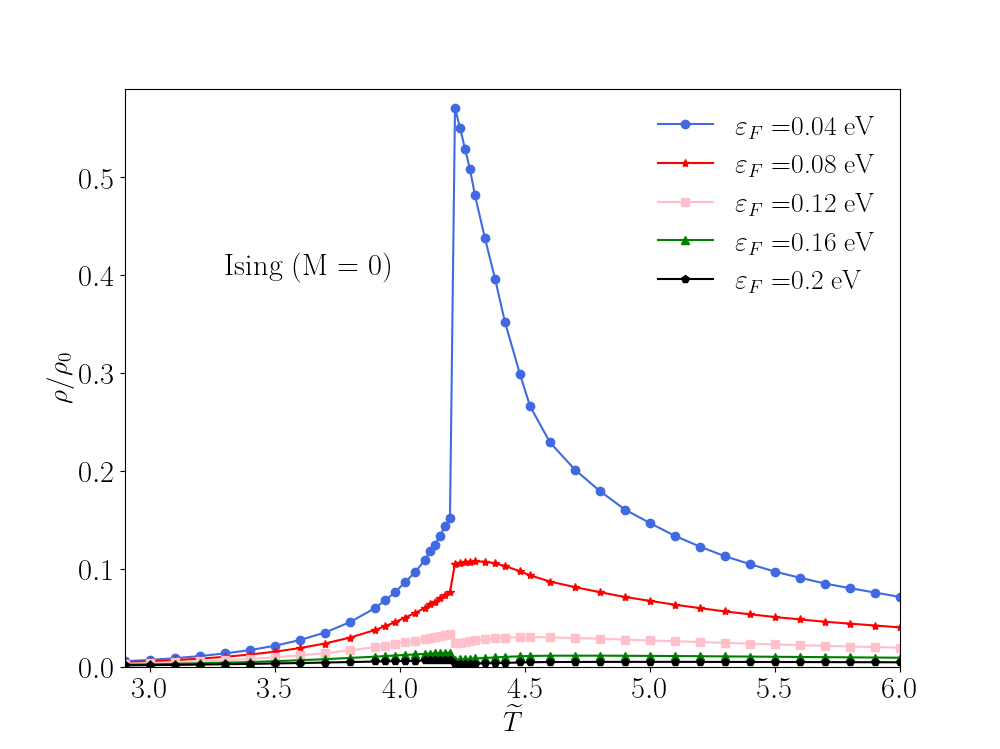}
	\includegraphics[width=0.5\textwidth]{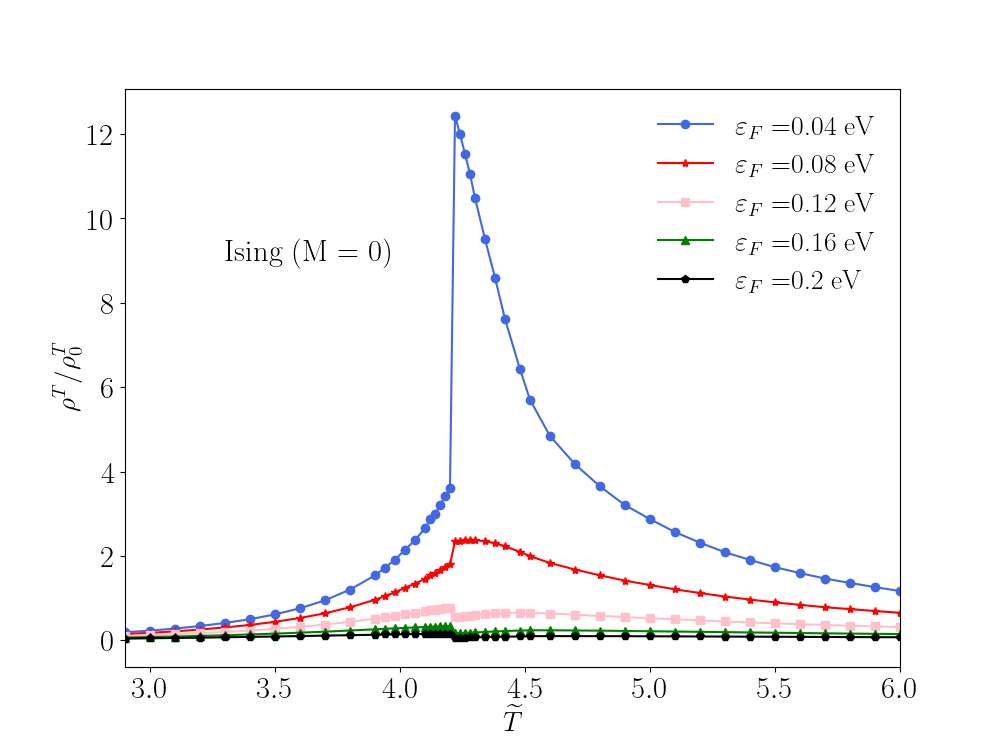}
	\caption{Transport properties of the antiferromagnetic TI surface as functions of the scaled temperature $\widetilde{T} = k_{\rm B} T/(J_1 \tilde{s}^2)$ for various Fermi energies. The magnetic interactions on the surface are modeled using the Ising Hamiltonian with NN exchange interaction $J_1$, where we set $\tilde{s} J_1 = 0.3$ meV, consistent with parameters for manganese bismuth telluride. The plots show: electrical resistivity (up) and thermal resistivity (down). Here, the interlayer interaction $J_c$ is scaled to $J_1$, and set to $-0.18$. The parameters $\rho_0$ and $\rho_0^T$ are defined as follows: $\rho_0 = \frac{4\pi^2 J_0^2 s^2 a^2}{\sqrt{3} \hbar^{2} v_{\rm F}^2} \frac{h}{e^2}$, $
		\rho_0^T = \frac{8\pi^3 J_0^2 s^2 a^3}{\sqrt{3} \hbar^2 v_{\rm F}^3 k_{\rm B}}$. 
		Both $\rho_0$ and $\rho_0^T$ are constants with dimensions of electrical resistivity and thermal resistivity, respectively. For magnetic TIs, with parameters $J_0=25~{\rm meV}$, $v_{\rm F}=5.5 \times 10^5 {\rm m/s}$, $a=4.33~\text{\AA}$, and $s=\frac{5}{2}$, these constants are estimated to be \(3250~\Omega\) and \(4.5\times 10^{7}~{\rm K/W}\), respectively.}
	\label{fig: r-rt-I}
\end{figure}

Figure \ref{fig: r-rt-I} presents the plots of surface electrical and thermal resistivities as functions of rescaled temperature $\widetilde{T}=k_{\rm B}T/(J_1\tilde{s}^2)$. Both resistivities exhibit similar behavior: At temperatures significantly below the surface critical temperature ($\widetilde{T}<4.2$), they remain small and nearly independent of the Fermi energy. As the temperature rises, the resistivities increase gradually. Near the critical temperature, they escalate sharply due to the influence of magnetic clustering on the surface of the TI.  Above the critical temperature, in the paramagentic phase, the resistivities decrease with increasing temperature, which are attributed to the behavior of magnetic clusters. At high temperatures, magnetic clusters shrink to the size of individual spins, and their number decreases, leading to a reduction in resistivity. 

At the critical temperature, both the electrical and thermal resistivities exhibit distinct behaviors across different Fermi energy values. Notable deviations in their temperature dependence emerge between low and high Fermi energies. Specifically, at a low Fermi energy of \(\ep_{\rm F} = 0.04\) meV, the resistivities show a pronounced increase at the critical temperature, followed by a gradual decline as temperature rises. In contrast, at a higher Fermi energy of \(\ep_{\rm F} = 0.12\) meV, the resistivities undergo a discontinuous drop at the critical temperature, then increase with temperature to form a broad maximum before gradually decreasing at elevated temperatures. These contrasting behaviors are attributed to variations in the relaxation time, governed by distinct dominant mechanisms: at low Fermi energies, the resistivities are primarily affected by the size of clusters, whereas at higher Fermi energies, the number of clusters play a more significant role in determining the relaxation dynamics.

\begin{figure}
	\centering
	\includegraphics[width=0.5\textwidth]{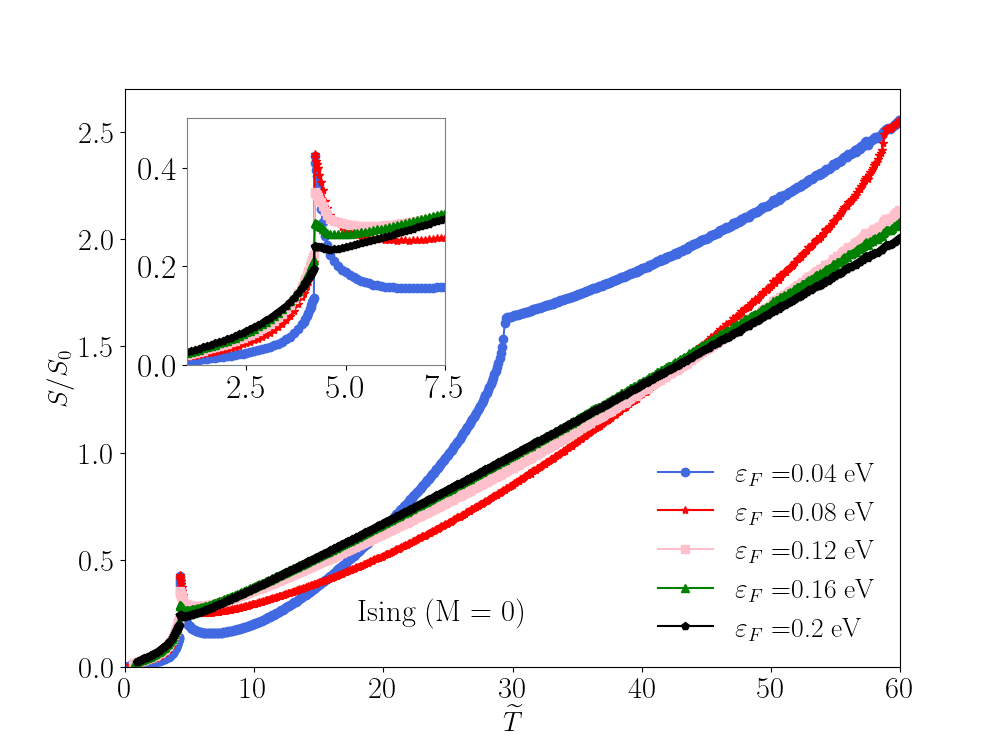}
	\includegraphics[width=0.5\textwidth]{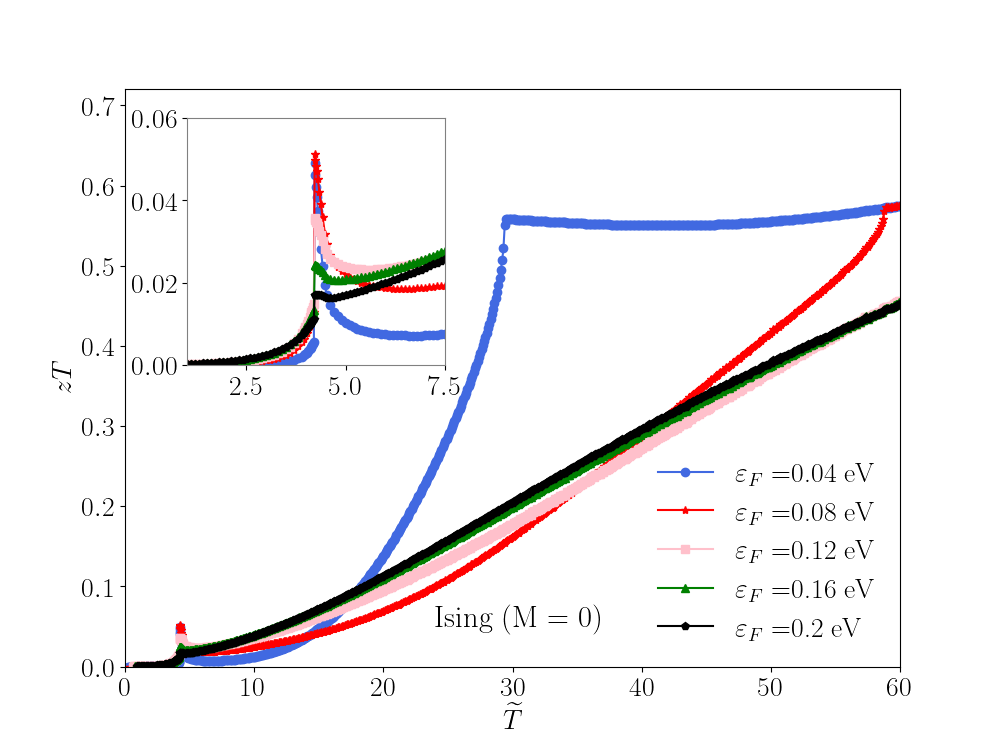}
	\caption{Thermoelectric properties of the surface of magnetic TIs as functions of $\widetilde{T}=k_{\rm B}T/(J_1\tilde{s}^2)$ for various Fermi energies. The magnetic interactions on the surface are modeled using the Ising Hamiltonian with NN exchange interaction $J_1$, where we set $\tilde{s} J_1 = 0.3$ meV, consistent with parameters for manganese bismuth telluride. The plots illustrate the thermopower (up) and the thermoelectric figure of merit (down), with the interlayer interaction $J_c$ is scaled to $J_1$, and set to $-0.18$. The thermopower is scaled to $S_0=\frac{k_{\rm B}}{e}\simeq 86.17 {\rm \mu V/K}$. Insets are plots that show the behavior near the critical temperature.}
	\label{fig: s-zT-I}
\end{figure}
In Fig. \ref{fig: s-zT-I}, the thermopower is presented as a function of temperature from 0 K to 522 K, corresponding to a scaled temperature $\widetilde{T}=60$. Manganese bismuth telluride compounds remain stable and retain their crystal structure within this temperature range. A maximum thermopower of approximately 215 $\mu \text{V}/\text{K}$ is observed at elevated temperatures well within the disordered phase, away from the critical point. This value is comparable to that of ${\rm Bi_2Te_3}$ and its alloys, indicating similar thermoelectric performance in this regime. In bismuth telluride alloys, such high thermopower is attributed to the widening of the energy gap and the suppression of intrinsic conduction effects \cite{Tang2015}. Conversely, in magnetic TIs, significant thermopower is primarily due to magnetic clustering on the surface. The measured value is also comparable to that of antiferromagnetic MnTe, a material used in spin-based thermoelectrics. In MnTe, paramagnon drag causes an anomalous increase in thermopower above the Neel temperature ($T_{\rm N} \simeq 305~\text{K}$), with reported values ranging from 100 to 200 $\mu \text{V}/\text{K}$ for pure to Li-doped MnTe \cite{Zheng2019}.

Across various Fermi energy values, two distinct features are observed in the behavior of both the thermopower and the thermoelectric figure of merit (see Fig. \ref{fig: s-zT-I}). One feature occurs at the critical point of the ferromagnetic–paramagnetic phase transition, while the other manifests at the temperature where the chemical potential vanishes.

At low temperatures, the thermopower and figure of merit are minimal due to the absence of magnetic clusters on the surface. As the temperature increases, magnetic clusters form, leading to a significant rise in thermopower. Near the surface critical temperature $\widetilde{T}_c \simeq 4.2$ (corresponding to $T_c \simeq 36.55~\text{K}$ for $\tilde{s}J_1 = 0.3$ meV), the thermopower exhibits an abrupt jump to a maximum. For higher Fermi energies, this peak is approximately $S \simeq 0.2 S_0 = 17.5~\mu\text{V}/\text{K}$, while for lower Fermi energies, it reaches $S \simeq 0.5 S_0 = 35~\mu\text{V}/\text{K}$.

Above $T_c$, magnetic clusters begin to break down and shrink to microscopic scales. In this regime, for low Fermi energies, the thermopower initially decreases but then begins to increase again, with a notable change in behavior around \(k_{\rm B} T \approx \sqrt{3}\E_{\rm F} / \pi \approx \E_{\rm F} / 2\). Interestingly, this rise occurs even as magnetic clusters continue to diminish. For high Fermi energies, the post-$T_c$ decrease is less pronounced, and the thermopower increases approximately linearly with temperature.

This behavior is explained by the temperature dependence of the chemical potential \(\mu(T)\) (see Eq. \ref{eq: mu}). When \(k_{\rm B} T = \sqrt{3}\E_{\rm F} / \pi\), \(\mu\) approaches zero. Combined with the linear energy dependence of the surface density of states in TIs, this leads to the observed anomaly in the thermopower.

Mathematically, this is reflected in the expression for thermopower derived from Eq. (\ref{eq: gamma}): $S = \gamma^{(1)}/(\gamma^{(0)}T)$. Here, $\gamma^{(1)}$ depends on $\mu$, while $\gamma^{(0)}$ does not. Consequently, abrupt changes in $\mu$ directly influence $\gamma^{(1)}$, resulting in non-monotonic features in the thermopower.

Finally, it is important to emphasize that magnetic clustering is the primary driver of the observed thermopower. The increase at higher temperatures disappears when interactions between magnetic atoms are neglected, underscoring the essential role of clustering.

\begin{figure}[h]
	\centering
	\includegraphics[width=0.5\textwidth]{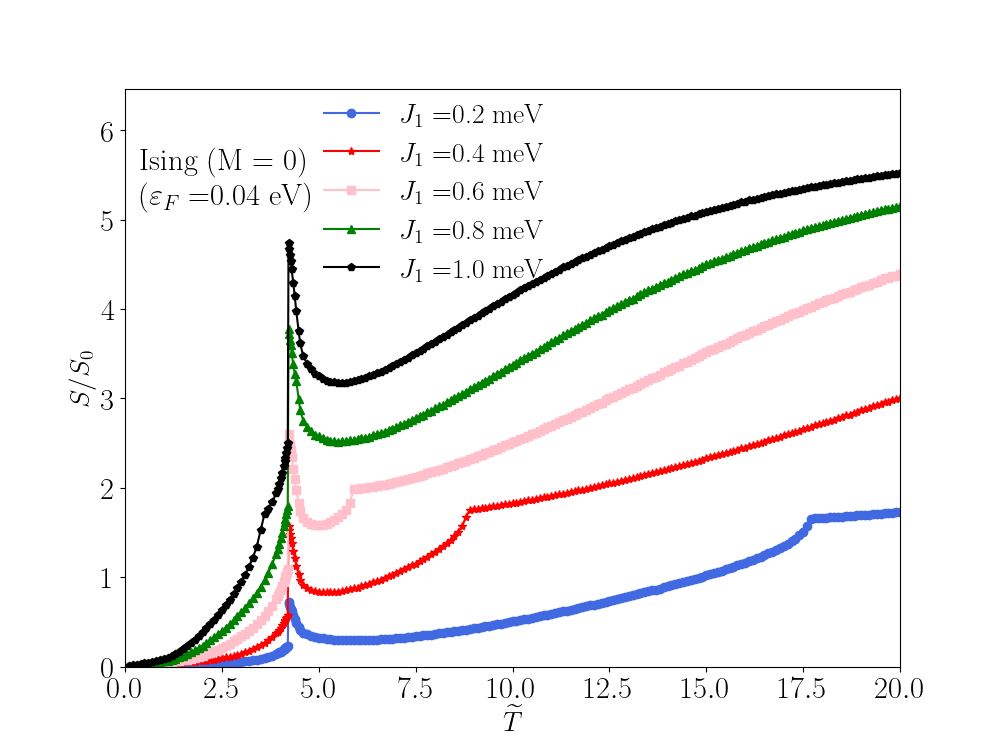}
	\includegraphics[width=0.5\textwidth]{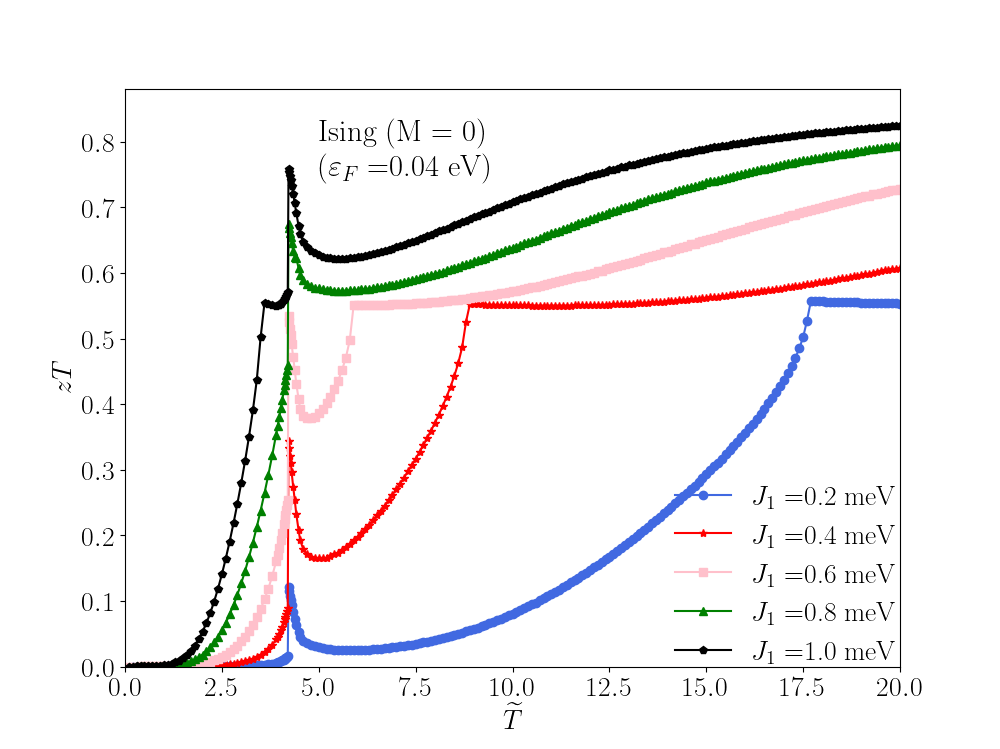}
	\caption{Thermoelectric properties of the surface of magnetic TIs as functions of $\widetilde{T}=k_{\rm B}T/(J_1\tilde{s}^2)$ for various $J_1$, at the Fermi energy $\ep_{\rm F}=0.04$ eV. The interaction between magnetic atoms on the TI surface is modeled using the Ising Hamiltonian with NN exchange interaction $J_1$. The interlayer interaction $J_c$ is scaled to $J_1$, and is fixed at $-0.18$ for all values of $J_1$. The plots illustrate the thermopower (up) and the thermoelectric figure of merit (down). The thermopower is scaled to $S_0=\frac{k_{\rm B}}{e}\simeq 86.17 {\rm \mu V/K}$.}
	\label{fig: s-zT-I-2}
\end{figure}

The thermopower anomaly is observed at temperatures from $~232$ ${\rm K}$ to 1160 ${\rm K}$, corresponding to Fermi energies from $40~\text{meV}$ to $200~\text{meV}$. The van der Waals materials such as $\rm MnBi_2Te_4$ remain stable up to approximately $670~\text{K}$, beyond which they decompose into $\rm MnTe$ and $\rm Bi_2Te_3$. This decomposition affects the material's magnetic properties, requiring adjustments to our results to assess thermoelectric properties at elevated temperatures. For $\rm MnTe$, a similar approach can be used to determine its thermoelectric properties.
Applying pressure increases the stability temperature of these materials. Under pressure, the magnetic properties remain largely unchanged, allowing us to continue using our results \cite{Htemperature1, Htemperature2}.

To investigate the impact of exchange couplings between magnetic atoms on thermoelectric properties, we plotted in Fig. \ref{fig: s-zT-I-2} the thermopower and figure of merit as functions of the scaled temperature \(\widetilde{T}\) for various strengths of the NN Ising interaction $J_1$. Although the qualitative behavior remains similar across all \(J_1\) values, featuring a peak at the critical temperature $k_{\rm B}T_c=6.25 J_1$ and a change in behavior at temperature where \(\mu=0\), the magnitudes of these quantities increase with larger \(J_1\).

Remarkably, the thermopower at $T_c$ exhibits a significant enhancement with increasing \(J_1\), reaching values not typically observed in standard thermoelectric materials. For instance, in systems with $J_1=1~\text{meV}$, at the critical temperature $k_{\rm B} T_c \approx 26.25~\text{meV} \sim 305~\text{K}$, the thermopower reaches approximately $S \sim 5 S_0 \sim 430~\mu\text{V/K}$, which is notably high compared to conventional thermoelectric materials. As a comparison, Li-dopped MnTe has a Neel temperature around 307 K with a magnon-drag thermopower of about 200 $\mu\text{V/K}$ at that temperature \cite{Zheng2019}. In contrast, at the surface of magnetic TIs, the thermopower exceeds this value by at least a factor of two.

Although thermopower continues to rise at higher temperatures, the integrity of the crystal structure becomes compromised at such elevated conditions, limiting the feasibility of practical applications.

\begin{figure}
	\centering
\includegraphics[width=0.5\textwidth]{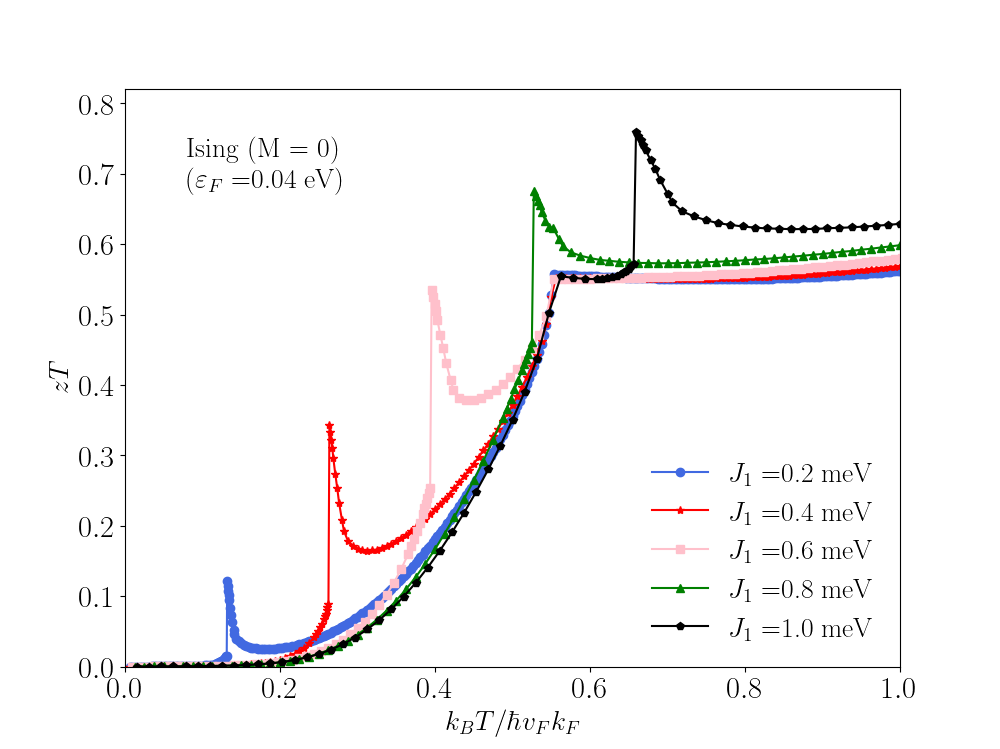}
	\caption{Thermoelectric figure of merit as a function of $k_{\rm B}T/\hbar v_{\rm F}k_{\rm F}$ for various NN exchange couplings $J_1$. Here, the interaction between magnetic atoms on the surface of the TI is governed by the Ising Hamiltonian. The antiferromagentic interlayer exchange $J_c$ is scaled to $J_1$, and set to $-0.18$. }
	\label{fig: ztH-Anti}
\end{figure}

In Fig. \ref{fig: ztH-Anti}, we present the figure of merit as a function of the normalized temperature \(k_{\rm B}T/\ep_{\rm F}\) for various strengths of \(J_1\).  The figure of merit exhibits two prominent changes in behavior at distinct temperatures. The first occurs at the critical temperature, which increases with larger \(J_1\). The second change occurs at the temperature where the chemical potential \(\mu\) equals zero; notably, this temperature is independent of the exchange coupling strength. Depending on the value of \(J_1\), the critical temperature may lie either below or above the temperature at which \(\mu = 0\). Beyond these transitions, the figure of merit gradually saturates, reaching a plateau around 0.8.  

\subsection{Ising case: Ferromagnetic interlayer interactions}

\begin{figure}
	\centering
	\includegraphics[width=0.5\textwidth]{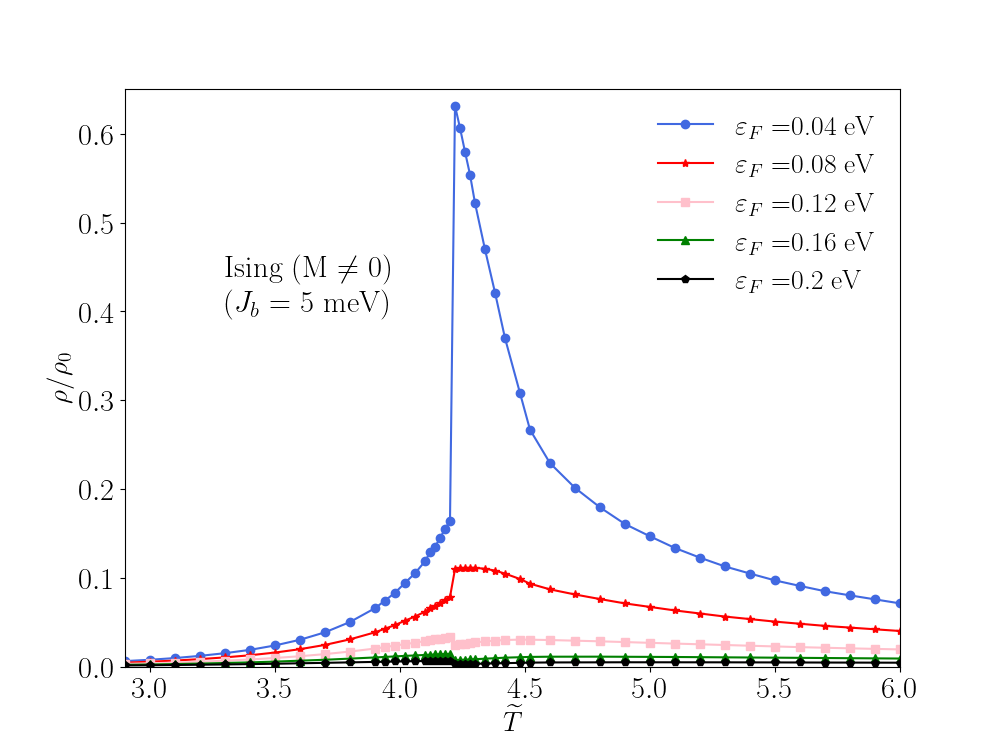}
	\includegraphics[width=0.5\textwidth]{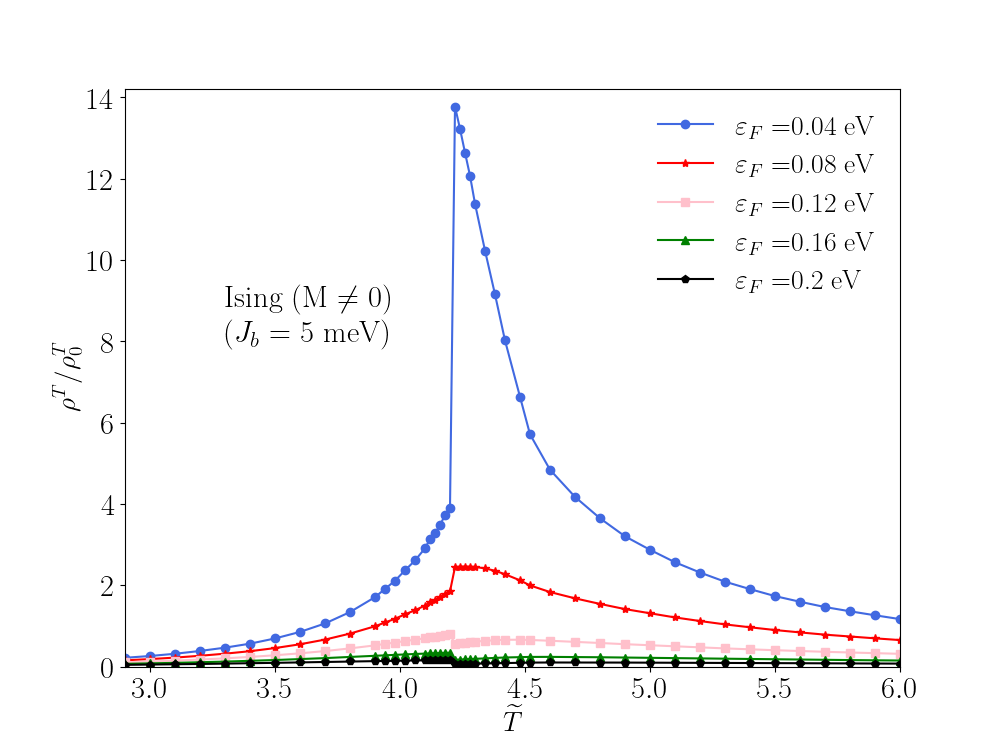}
	\caption{Transport properties of the surface of the magnetic TI as functions of $\widetilde{T}=k_{\rm B}T/(J_1\tilde{s}^2)$, for various Fermi energies. The magnetic interactions on the surface are modeled using the Ising Hamiltonian with NN exchange interaction $J_1$, where we set $\tilde{s} J_1 = 0.3$ meV, consistent with parameters for manganese bismuth telluride. Here, the ferromagentic interlayer exchange $J_c$ is scaled to $J_1$, and set to $+0.18$. The plots show: electrical resistivity (up) and thermal resistivity (down). The constants $\rho_0$ and $\rho_0^T$ are defined in the caption of Figure \ref{fig: r-rt-I}.}
	\label{fig: r-rT-IFerro}
\end{figure}
\begin{figure}
	\centering
	\includegraphics[width=0.5\textwidth]{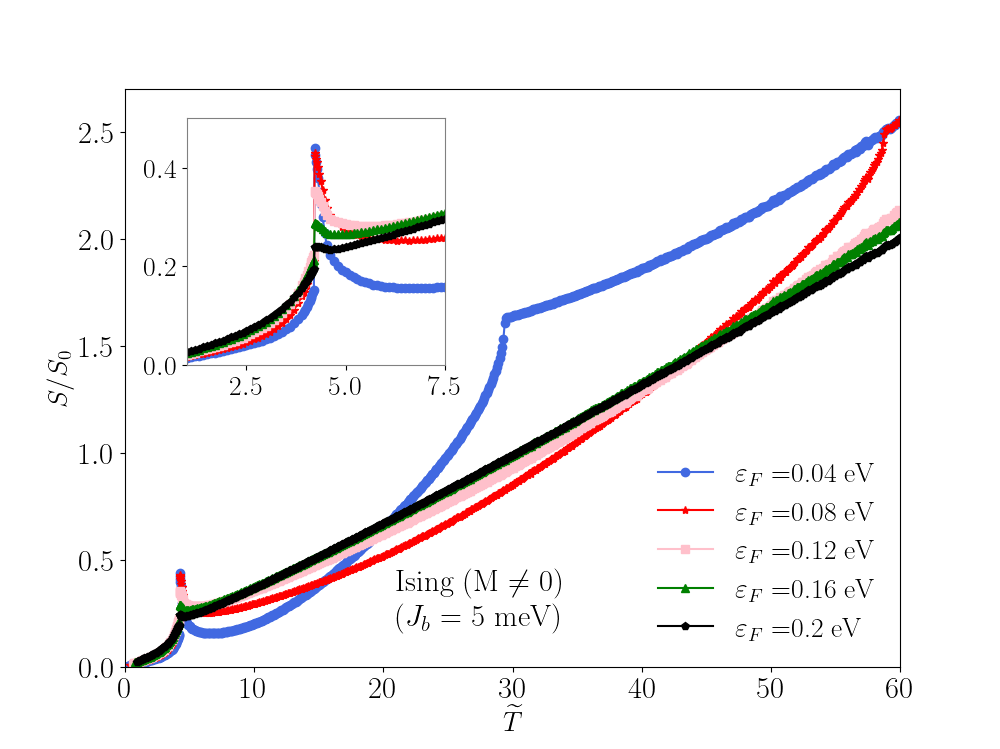}
	\includegraphics[width=0.5\textwidth]{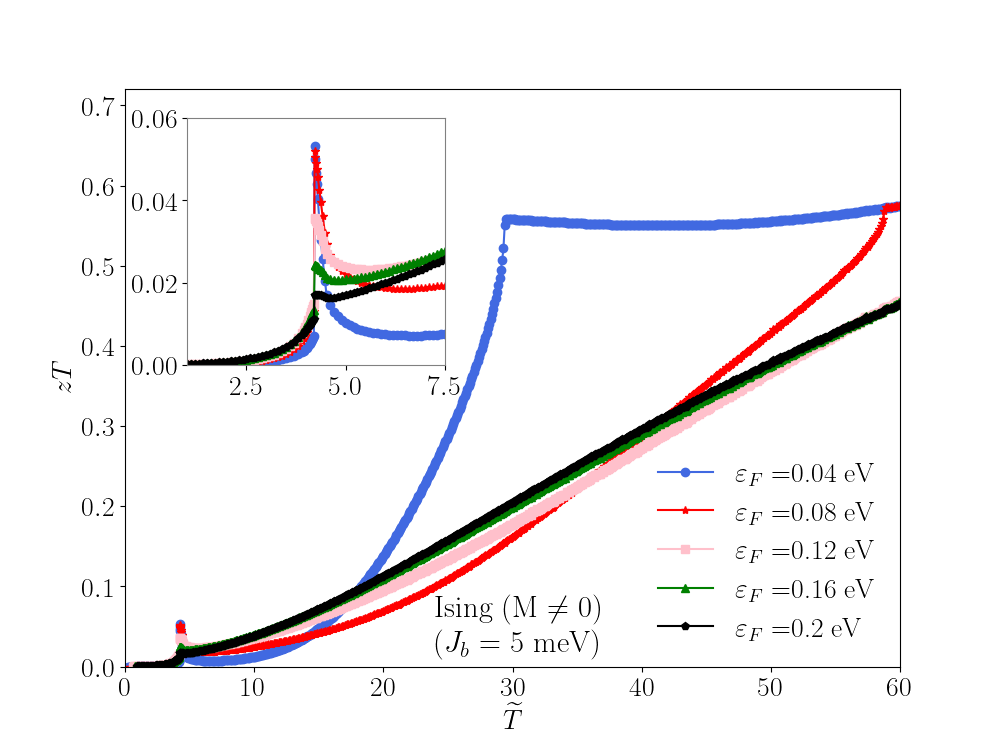}
	\caption{Thermoelectric properties of the surface of the magnetic TI as functions of $\widetilde{T}=k_{\rm B}T/(J_1\tilde{s}^2)$. The magnetic interactions on the surface are modeled using the Ising Hamiltonian with NN exchange interaction $J_1$, where we set $\tilde{s} J_1 = 0.3$ meV, consistent with parameters for manganese bismuth telluride. Here, the ferromagentic interlayer exchange $J_c$ is scaled to $J_1$, and set to $+0.18$. The plots show: thermopower (up), and thermoelectric figure of merit (down). Here, $S_0=\frac{k_{\rm B}}{e}\simeq 86.17 \mu \text{V}/\text{K}$. Insets are plots that show the behavior near the critical temperature.
	}
	\label{fig: s-zT-IFerro}
\end{figure}
For intrinsic ferromagnetic TIs like \(\rm MnBi_6Te_{10}\), the bulk exhibits temperature-dependent magnetization \(M(T)\), which in turn induces a temperature-dependent energy gap in the surface electrons. Below the bulk critical temperature (\(T<T_c^{\text{bulk}}\)), the surface states possess a finite energy gap. However, above this critical temperature, the gap closes. This behavior significantly influences the thermoelectric properties of the TI, as detailed below.

The temperature dependence of electrical and thermal resistivities (see Fig. \ref{fig: r-rT-IFerro}), as well as thermopower and the figure of merit (see Fig. \ref{fig: s-zT-IFerro}), are generally similar for both antiferromagnetic and ferromagnetic interlayer interactions. However, there is a subtle difference in thermopower. Below the bulk critical temperature ($T_c^{bulk}$), the bulk exhibits spontaneous magnetization, leading to the emergence of a surface energy gap. Within this temperature range, significant variations are observed in thermopower and the figure of merit, primarily attributable to the presence of the energy gap, which enhances thermopower below $T_c^{bulk}$. Conversely, above $T_c^{bulk}$, the bulk magnetization vanishes, resulting in thermoelectric properties that closely resemble those of an antiferromagnetic bulk, where the bulk magnetization is inherently zero. 
This phenomenon occurs because interlayer interactions, besides altering the surface gap, also influence the size and number of clusters. Ferromagnetic interlayer interactions can modify the clusters in a way that leads to thermopower saturation.
It is important to note that the use of an Ising interaction model to describe the bulk magnetization ensures that the surface critical temperature closely aligns with the bulk critical temperature.

%%%%%%%%%%%%%%%%%%%%%%%%%%%%%
\subsection{Heisenberg interaction between magnetic atoms on the surface of the TI}

\begin{figure}
	\centering
	\includegraphics[width=0.5\textwidth]{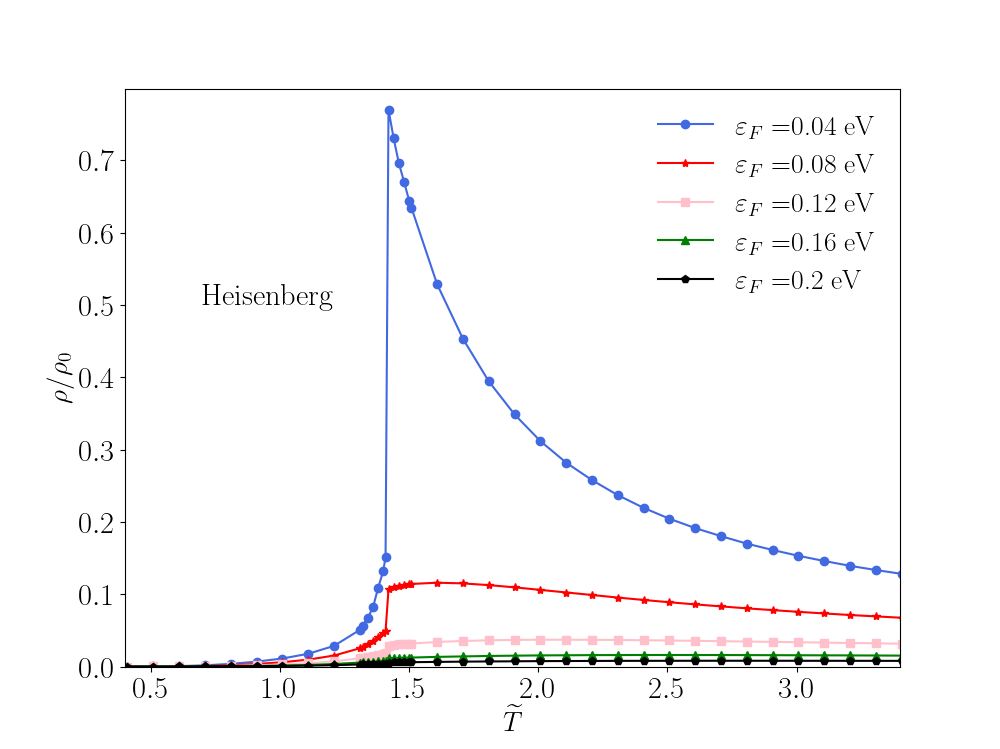}
	\includegraphics[width=0.5\textwidth]{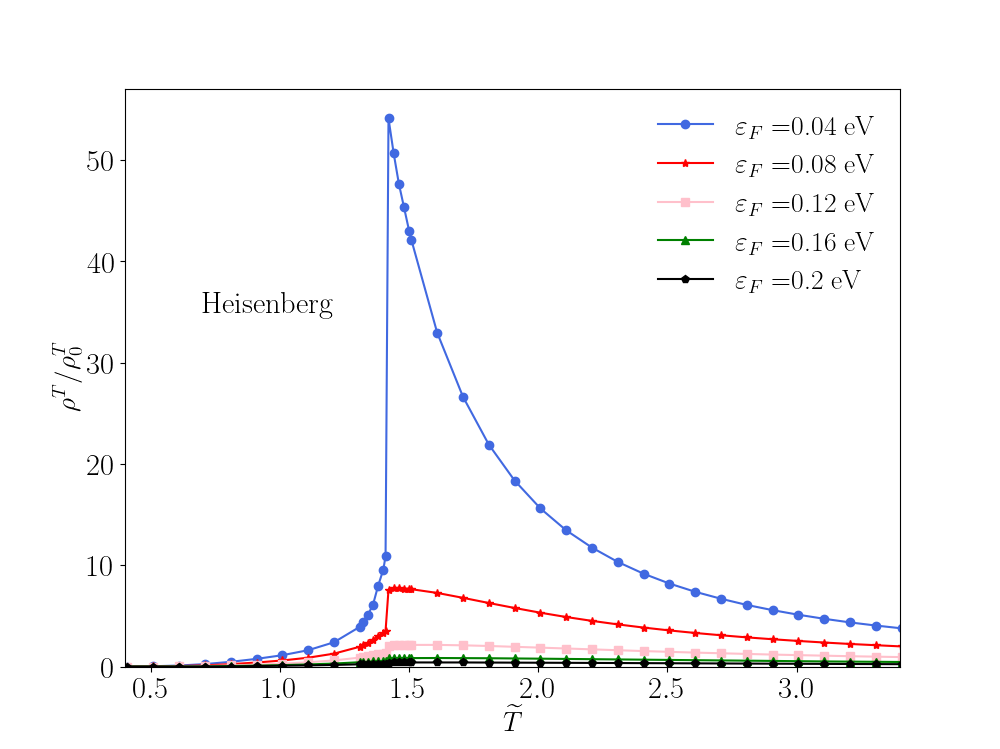}
	\caption{Transport properties of the surface of the magnetic TI as functions of $\widetilde{T}=k_{\rm B}T/(J_1\tilde{s}^2)$ for various Fermi energies. The interaction between magnetic atoms on the surface of the TI is described by the Heisenberg Hamiltonian in Eq. (\ref{eq:H1}). Here, $J_1$ represents the NN exchange coupling, while $J_2$, $J_4$, the uniaxial single ion anisotropy $D$, and the antiferromagnetic interlayer exchange $J_c$ are scaled relative to $J_1$, with values set to $-0.28$, $0.08$, $0.4$, and $-0.18$, respectively. Here, we set $\tilde{s} J_1 = 0.3$ meV, consistent with parameters for manganese bismuth telluride. The plots show: electrical resistivity (up) and thermal resistivity (down). The constants $\rho_0$ and $\rho_0^T$ are defined in the caption of Figure \ref{fig: r-rt-I}.}
	\label{fig: r-rT-HAnti}
\end{figure}

When the interaction between magnetic atoms on the surface of magnetic TIs is described by the Heisenbeg model in Eq. (\ref{eq:H1}), the relaxation time of Dirac electrons is given by Eq. (\ref{eq: tau2}). For intrinsic antiferromagnetic TIs, the bulk magnetization $M$ is zero for all temperatures, resulting in a vanishing surface gap. Consequently, $\cos\eta_k=0$, and the relaxation time simplifies to:
\begin{equation}
	\label{eq: tau22}
	\frac{1}{\tau_{k,\theta_s}} =\frac{1}{\tau_0}\bar{n}_{\theta_s} k \xi_{\theta_s} \left(\frac{\xi_{\theta_s}}{a}\right)^3\frac{2+\cos^2\theta_s+2k^2\xi_{\theta_s}^2\sin^2\theta_s}{(1+4k^2\xi_{\theta_s}^2)^{5/2}}.
\end{equation}
By utilizing Eq. (\ref{eq: tau22}) and numerically solving the integral in Eq. (\ref{eq: gamma}), we obtain the electrical and thermal resistivities as functions of temperature, as illustrated in Fig. \ref{fig: r-rT-HAnti}.
Overall, while the resistivity behaviors in the Heisenberg model bear a strong resemblance to those in the Ising case, several key differences stand out. At low temperatures, both the electrical and thermal resistivities approach zero, largely independent of the Fermi energy. However, at the critical temperature $\widetilde{T}_c=1.42$ (corresponding to $T_c \simeq 12.3~\text{K}$ for manganese bismuth telluride with $\tilde{s}J_1 = 0.3$ meV), the Heisenberg model displays a discontinuous increase in resistivities across all Fermi energy values. This contrasts sharply with the Ising model, which exhibits an abrupt decrease at higher Fermi energies. Additionally, the peak thermal resistivity at the critical point is markedly higher in the Heisenberg model, exceeding that of the Ising model by several times.

These differences in low-temperature and critical behavior stem from the distinct magnetic clustering mechanisms inherent to the Ising and Heisenberg models. Above the critical temperature, however, their resistivity profiles converge, with both models exhibiting a gradual decline as the temperature increases.

The weak surface resistivity observed in the ferromagnetic phase of the surface pertains only to surface properties. In contrast, the bulk electrical resistivity in the antiferromagnetic (AFM) phase (the magnetic order of the bulk of the antiferromagnetic TI) behaves differently \cite{experiment_2019, experiment_2020}. Experimental evidence shows that in an A-type AFM system, interlayer AFM coupling typically causes strong spin scattering in the ordered state, leading to high resistivity. However, when an external magnetic field is strong enough to overcome this coupling, spin scattering is suppressed, resulting in low resistivity. This phenomenon, known as the spin-valve effect, was first demonstrated in magnetic multilayer thin films. The spin-valve effect can also occur between an A-type AFM and a canted AFM state, as observed in \(\rm Ca_3Ru_2O_7\) and \(\rm MnBi_2Te_4\) during magnetotransport measurements \cite{experiment_2019}.

\begin{figure}
	\centering
	\includegraphics[width=0.5\textwidth]{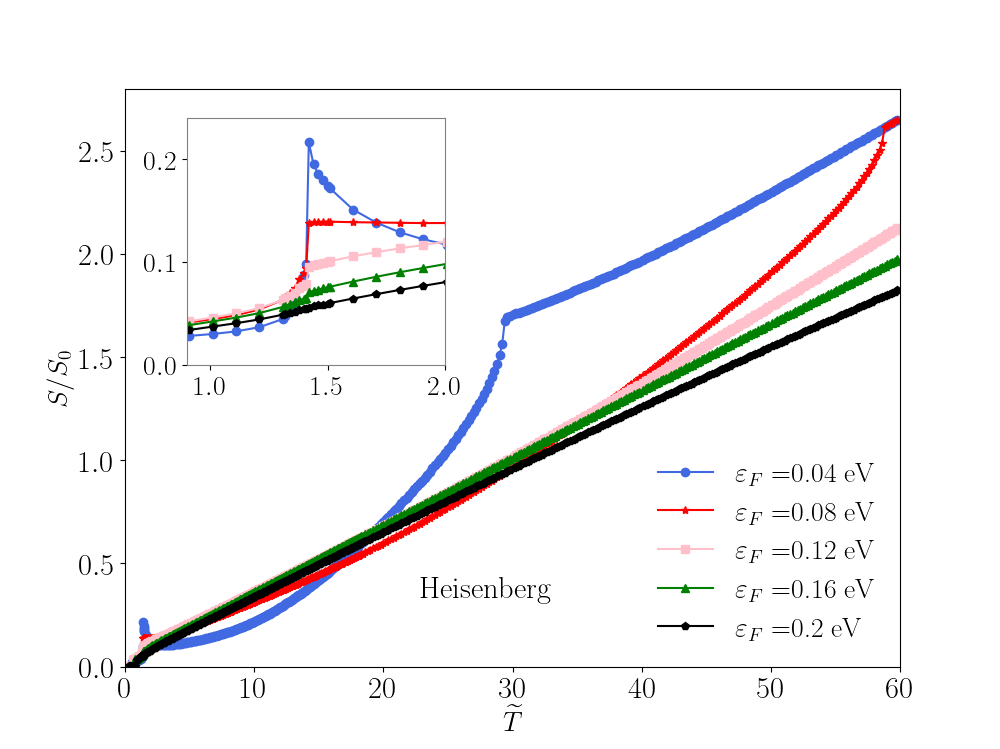}
	\includegraphics[width=0.5\textwidth]{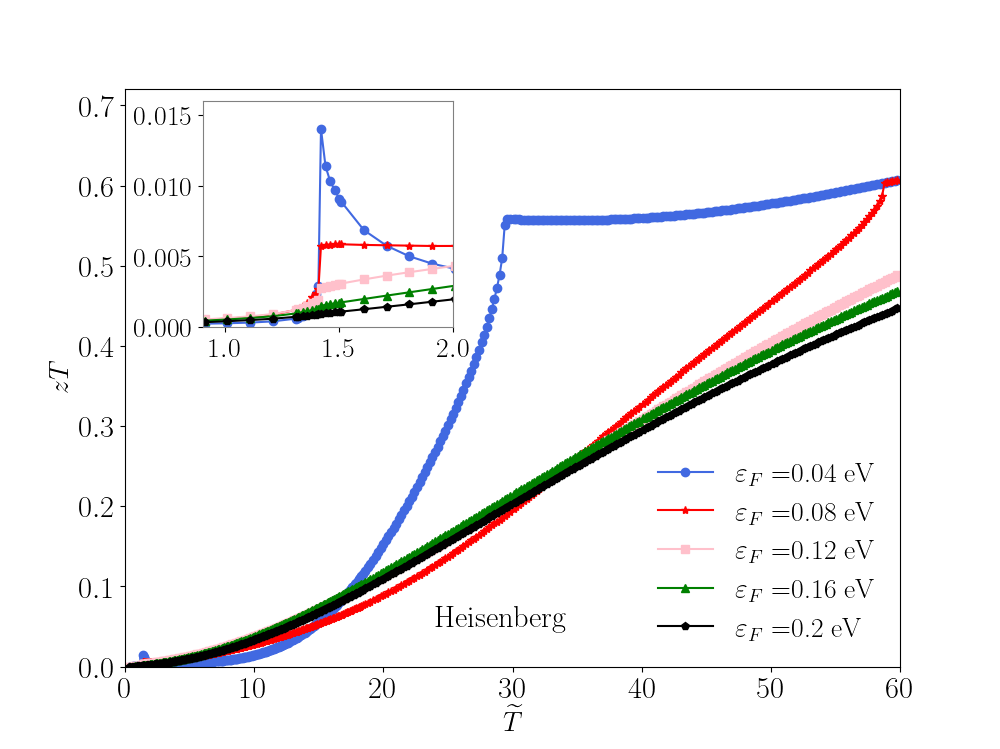}
	\caption{Thermoelectric properties of the surface of the magnetic TI as functions of temperature $\widetilde{T}=k_{\rm B}T/(J_1\tilde{s}^2)$, for various Fermi energies. The interaction between magnetic atoms on the surface of the TI is described by the Heisenberg Hamiltonian in Eq. (\ref{eq:H1}). Here, $J_1$ represents the NN exchange coupling, while $J_2$, $J_4$, the uniaxial single ion anisotropy $D$, and the antiferromagnetic interlayer exchange $J_c$ are scaled to $J_1$, with values set to $-0.28$, $0.08$, $0.4$, and $-0.18$, respectively. Here, we set $\tilde{s} J_1 = 0.3$ meV, consistent with parameters for manganese bismuth telluride. The plots show: thermopower (up), and thermoelectric figure of merit (down). Here, the interlayer interaction is antiferromagnetic and set to $J_c/J_1 = -0.18$.
		% and $S_0=\frac{k_{\rm B}}{e}\simeq 86.17 \mu \text{V}/\text{K}$.
	}
	\label{fig: s-zT-HAnti}
\end{figure}
The temperature dependence of thermopower in the Heisenberg case is depicted in Fig. \ref{fig: s-zT-HAnti}. When intra-layer interactions are described by the Heisenberg Hamiltonian, the surface of the TI transitions to the paramagnetic phase at a significantly lower temperature than observed in the Ising model. At the critical point, the thermopower displays a local peak similar to that in the Ising model; however, this peak is roughly half the magnitude of its Ising counterpart. In the ordered ferromagnetic phase, the thermopower remains very low, while in the disordered phase, it experiences an anomalous jump and increases with temperature. This behavior is largely attributed to magnetic clustering, where, in the paramagnetic phase, the size and number of magnetic clusters are greater than those in the ferromagnetic phase.

The relaxation time fundamentally dictates the behavior of the electrical resistivity and, consequently, the thermal resistivity. However, the thermopower ($S$) and the thermoelectric figure of merit ($zT$) are governed by additional physical parameters beyond the relaxation time, which explains their different behavior.

To clarify this distinction, we refer to the Onsager expression for the thermopower, given by $S = \gamma^{(1)} / (T \gamma^{(0)})$, where according to Eq. (\ref{eq: gamma}), $\gamma^{(1)}$ has a distinct dependence on the chemical potential $\mu$, while $\gamma^{(0)}$ (which is proportional to the electrical conductivity) is unaffected by the sign of $\mu$. Therefore, the thermal evolution of $\mu$ directly influences $\gamma^{(1)}$, and its interplay with the relaxation time results in the unique behavior of the thermopower not observed in the electrical resistivity.

While the thermal resistivity also depends on the chemical potential, its behavior generally follows a similar dependence on the relaxation time as the electrical resistivity. To isolate the thermodynamic functions that primarily govern the thermopower, we can consider an energy-independent relaxation time. Although this is not the case in our system, this assumption helps identify which physical quantities, besides the relaxation time, are vital. Under this assumption, the thermopower at a given temperature simplifies to:
\begin{equation}
S=\frac{1}{eT}\frac{\int D(\E)d\E(-\frac{\partial f^0}{\partial\E})\E}{\int D(\E)d\E(-\frac{\partial f^0}{\partial\E})}-\frac{\mu}{eT}.
\end{equation}
This expression can be rewritten as \cite{Jazandari2025}:
\begin{equation}
S =\frac{1}{eT} \left( \frac{ \frac{dU}{d\mu} - \mu \frac{dN}{d\mu} }{ \frac{dN}{d\mu} } \right)_{T}=\frac{1}{eT} \frac{ T \left( \frac{\partial {\cal S}}{\partial \mu} \right)_{T} }{ \left( \frac{\partial N}{\partial \mu} \right)_{T} } = \frac{1}{e} \left( \frac{\partial {\cal S}}{\partial N} \right)_{T}
\label{eq: S-Se}
\end{equation}
where $U$ is the internal energy, $N$ is the number of carriers, and ${\cal S}$ is the entropy. The Eq. (\ref{eq: S-Se}) indicates that the entropy per carrier plays an essential role in determining the thermopower, in addition to the relaxation time.

In the general case of an energy-dependent relaxation time, such as in our magnetic TIs, it has been shown that the magnon-drag thermopower arising from electron-magnetic atom interactions is given by \cite{Jazandari2025}:
\begin{equation}
S_{md}=\frac{1}{e}\frac{s_m}{n_e}(\frac{\tau_m}{\tau_m+\tau_{me}}),
\label{eq:mag-drag}
\end{equation}
where $s_m$ is the magnonic entropy and $n_e$ is carrier density. This expression depends not only on the relaxation times ($\tau_m$ for magnon scattering and $\tau_{me}$ for electron-magnon scattering) but also critically on the entropy per carrier. The relaxation time studied in our work, $\tau$, is the electron relaxation time due to magnetic clusters, which is proportional to $\tau_{me}$ by a factor related to temperature and magnon group velocity \cite{Sugihara1972, Zanmarchi1968}. The Eq. (\ref{eq:mag-drag}) provides a clear physical explanation for why the thermopower does not simply mirror the behavior of $\tau$.

We have also illustrated the thermoelectric figure of merit as a function of temperature in Fig. \ref{fig: s-zT-HAnti}. Below the surface critical temperature, the figure of merit approaches zero. At the critical point, it begins to increase with a slight jump, ultimately reaching a saturation value of approximately $0.6$ at the temperature where the chemical potential becomes zero. For example, with a Fermi energy of $40~\text{meV}$, the saturation temperature is $230~\text{K}$. This result is particularly advantageous when compared to other thermoelectric materials that typically operate within mid-temperature ranges of $400$ to $900~\text{K}$ \cite{Tang2015}.

In magnetic systems, enhancing thermoelectric performance, specifically the thermopower and the thermoelectric figure of merit, is typically achieved through doping with electrons or holes. For example, in antiferromagnetic MnTe, doping with Li increases the thermopower to approximately 200 $\mu$V/K and raises the figure of merit above unity at temperatures exceeding 900 K. In our study, we attain comparable efficiency by tuning the exchange interaction $J_1$ between magnetic atoms. However, as we demonstrated, increasing $J_1$ and adjusting the Fermi energy significantly enhances the thermopower on the surface of the magnetic TI, yet the figure of merit remains below 1. Therefore, when compared to other materials, the magnetic TI surface exhibits relatively lower thermoelectric efficiency.

It is important, however, to compare these results at the same temperature. At room temperature, although the figure of merit for the magnetic TI does not exceed 1, it is nearly four times higher than that of Li-doped MnTe. For instance, at the critical temperature, Li-doped MnTe exhibits a figure of merit close to 0.1 \cite{Zheng2019}, whereas under similar conditions, specifically with $J_1=1$ meV and a Fermi energy of 40 meV, the magnetic TI achieves a value near 0.8.

%%%%%%%%%%%%%%%%%%%%%%%%%%%%%

\section{Comparison with experiments}

Extracting the surface resistivity-both electrical and thermal-from experimental data, and disentangling it from bulk contributions, presents substantial methodological challenges. This complexity hampers direct comparison between theoretical predictions and experimental measurements. Aside from specific cases, such as the direct measurement of the surface resistivity of \(\rm{Bi_2Te_2Se}\) TI in the presence of surface steps \cite{Ko2022}, the separation of surface and bulk resistivity contributions in magnetic TIs remains largely unresolved in laboratory studies, as confirmed by our literature review. Nonetheless, the surface resistivity behavior predicted by the magnetic clustering model aligns qualitatively with experimental findings. Notably, the resistivity value derived from the clustering approach-approximately 3250 $\Omega$ at temperatures slightly above the critical temperature (see Fig. \ref{fig: r-rT-HAnti})-is of the same order of magnitude as the surface resistivity reported in \cite{Ko2022} for \(\rm Bi_2Te_2Se\).

In \cite{Yan2019}, the authors reported the growth of millimeter-sized \(\rm MnBi_2Te_4\) single crystals and observed a drop in electrical resistivity upon cooling across the bulk transition temperature \(T_{\rm N}\). They noted critical scattering near \(T_{\rm N}\) in thermal conductivity, but thermopower showed a linear temperature dependence from 2 K to 300 K without anomalies around \(T_{\rm N}\).

In another study \cite{experiment_2019}, the authors systematically measured the variation of resistivity of \(\rm MnBi_2Te_4\) in the absence/presence of an external magnetic field. Their results show that the electrical resistivity in $x$ and $z$ directions, $\rho_{xx}(T)$ and $\rho_{zz}(T)$, peak at the Neel temperature, and the peaks are suppressed by the magnetic fields higher than a critical value. This observation is consistent with the spin fluctuation-driven spin scattering scenario. 

In \cite{experiment_2020}, the electrical transport measurements were carried out on a \(\rm MnBi_2Te_4\) single crystal with a thickness of $\sim 10\mu \text{m}$. Their results display the metallic temperature dependence of the longitudinal resistivity $\rho_{xx}$ from 1.5 K to room temperature. A sharp transition at $\sim 25 \text{K}$ corresponding to the AFM transition is consistent with $T_{\rm N}$ from the magnetic property measurement.

In \cite{experiment_2022}, it has been shown that a kink-like transition is observed around the Neel temperature in both the thermopwer and electrical resisitivity of ${\rm MnBi_2Te_4}$. Upon approaching $T_{\rm N}$, spin fluctuations are expected to be stronger, leading to stronger scattering and thus a shorter mean-free-path of electrons and an increase in resistivity. As the long-range magnetic order develops below $T_{\rm N}$, in which spins within the $x-y$ plane are ferromagnetically aligned, spin fluctuations and the electron-spin scattering are suppressed, resulting in kink in $\rho$ at $T_{\rm N}$ followed by a continuous decease upon decreasing the temperature. For thermopower, its negative sign over the whole
measured temperature range suggests electrons as the dominant charge carriers, which is consistent with the Hall effect transport studies. As the temperature decreases, the entropy per electron decreases, leading to a decrease in thermopower, on the other hand, as will be discussed later, below $T_{\rm N}$ the magnon-electron drag effect due to the coherent momentum conserving magnon-electron scattering gives rise to an increase in thermopower. As a result, the net thermopower develops a broad bump around 14 K. 
The maximum electron contribution below 25 K is 0.05 $\text{W}/\text{mK}$, much smaller than the increase of thermal conductivity observed below $T_{\rm N}$. Therefore, the main heat carriers in $\rm {MnBi_2Te_4}$ are phonons and/or magnons. Magnons can contribute to the thermal conduction via two ways. One is that magnons
serve as heat carriers, which would lead to an enhanced thermal conductivity below $T_{\rm N}$, the other is that magnons scatter phonons, giving rise to a reduction of the phonon contribution to thermal conductivity. Thus, the increase of thermal conductivity below $T_{\rm N}$ indicates the dominance of magnons as heat carriers.

\section{Summary and outlook}\label{sum}

Identifying and developing thermoelectric materials capable of converting waste heat into electricity is crucial for advancing clean energy technologies. Among the various materials studied, bismuth telluride and its alloys have received significant attention. This paper theoretically investigates the thermoelectric properties of the surface of magnetic TIs. Our findings demonstrate that the thermopower of the surface of magnetic TIs is comparable to other high-performance thermoelectric materials. The interaction of magnetic atoms on the surface plays a crucial role in achieving high thermopower. Specifically, the exchange interactions between magnetic atoms lead to the formation of magnetic clusters, which scatter Dirac electrons and significantly enhance the thermoelectric properties of the surface.

The results of this study are presented in detail below.

%%%%%%%%%%%%%%%%
\begin{enumerate}
	\item{{\it Magnetic clustering in Ising and Heisenberg models: A comparative analysis below and above surface critical temperature}
		
		Below the surface critical temperature, the definition of magnetic clusters in the Ising and Heisenberg models is analogous. In the Ising model, spin alignment serves as the criterion for cluster membership. Given that spins in the Ising model assume discrete values of $+1$ and $-1$, this definition is quite effective. However, in the Heisenberg model, spins are vectors with continuous orientations in space, making the alignment criterion less efficient. Therefore, we employ the Wolf probability for magnetic clustering.
		
		Above the critical temperature, within the disordered paramagnetic phase, clustering in the Ising model is based on the energy of links, whereas in the Heisenberg model, the Wolf probability is again utilized. The Wolf probability assigns a greater weight to neighboring spins that are aligned in the same direction. Consequently, on the surface of magnetic TIs with Heisenberg spin interaction, in the paramagentic phase, short-range ordered ferromagnetic domains are considered magnetic clusters. This approach is not applicable to the Ising model, where spins are restricted to two directions. In the Ising model, spins with the highest link energy with their neighbors are regarded as centers of magnetoresistance generation. For a spin to have the strongest link with its neighbors, it must be aligned with all of them.
		
We derive our definition of a cluster from GMR experiments. In these experiments, electrons with down spins are scattered by electrons with up spins, while those with up spins pass through without scattering. This concept is discussed in our paper. The initial direction that electrons on the surface of TIs choose is influenced by the initial magnetization, an effect we have implicitly accounted for. Consequently, we define clusters as those that align oppositely to the direction of magnetization. This idea is clearly articulated in the Ising model. However, in the Heisenberg model, where we have various spin directions, we apply the Wolf probability while still maintaining the constraint of opposite alignment to the magnetization.}
	
\item{{\it Inaccuracies in thermopower predictions by Mott relation in magnetic TIs}:

According to the Mott relation, the thermopower is generally expressed as \cite{Mahan}:
\begin{equation}
S = \frac{\pi^2 k_{\rm B}^2 T}{3 e} \left. \frac{d \ln \sigma(\E)}{d \E} \right|_{\E = \E_{\rm F}},
\end{equation}
where $\sigma(\E)$ is the electrical conductivity (see Eq. (\ref{eq: gamma})). At the surface of magnetic TIs, in the case of gapless Dirac electrons ($\E_k=\hbar v_{\rm F} k$), the Mott relation simplifies to:
\begin{equation}
S = \frac{\pi^{2} k_{B}^2 T}{3 e} \left. \frac{\partial (\tau_{k} \E_{k})}{\E_k\partial \E_k} \right|_{\E_{k} = \E_{\rm F}},
\label{eq: Mott}
\end{equation}
where $\tau_{k}$ is the relaxation time. 

In Fig. \ref{fig: mott}, we present the surface thermopower of magnetic TIs predicted by the Mott relation in Eq. (\ref{eq: Mott}) as a function of temperature.
\begin{figure}
	\centering
	\includegraphics[width=0.5\textwidth]{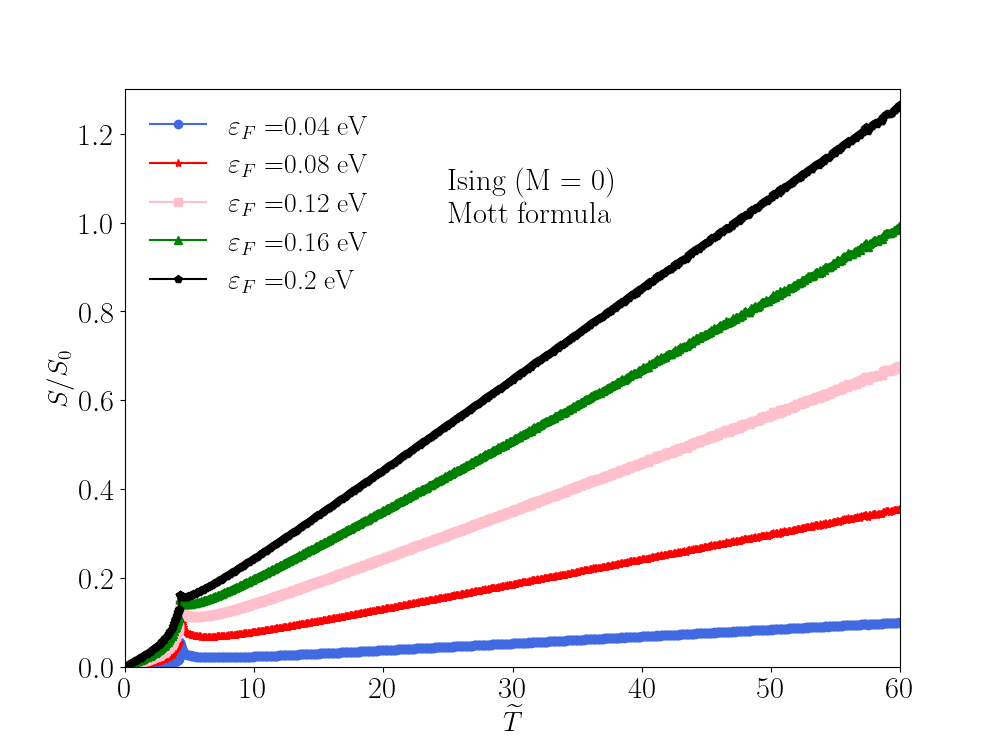}
	\caption{Surface thermopower of magnetic TIs as a function of the scaled temperature $\widetilde{T} =k_{\rm B} T/(J_1 \tilde{s}^2)$ for various Fermi energies, computed via the Mott relation. The magnetic atoms on the TI surface interact through an Ising Hamiltonian with NN exchange coupling $J_1$. The interlayer interaction $J_c$ is scaled relative to $J_1$ and set to $-0.18$. Here, we set $\tilde{s} J_1 = 0.3$ meV, consistent with parameters for manganese bismuth telluride. The thermopower is scaled to $S_0=\frac{k_{\rm B}}{e} \simeq 86.17~\mu\text{V/K}$.}
	\label{fig: mott}
\end{figure}
A comparison with our data in Fig. \ref{fig: s-zT-I} (upper panel) reveals that, at low temperatures, the temperature dependence of the Mott thermopower aligns closely with our findings. However, as the temperature rises, the Mott prediction begins to significantly diverge from our results.

Firstly, the change in behavior observed at the temperature where the chemical potential $\mu=0$ is not captured by the Mott thermopower. This discrepancy arises because, in the Mott relation, the chemical potential is approximated by the Fermi energy, making the conductivity effectively independent of the chemical potential.

Secondly, whereas the Mott relation indicates that, at temperatures above $T_c$, the thermopower increases with increasing Fermi energy, the thermopower predicted by the Onsager relation in Eq. (\ref{eq: gamma}) does not exhibit a uniform dependence on the Fermi energy.

In systems with a linear density of states, such as the surface states of magnetic TIs, the energy dependence of the conductivity becomes more pronounced and exhibits significant temperature dependence. Consequently, the assumptions underlying the Mott relation become less valid, and deviations from its predictions can occur. In such cases, it is necessary to employ the full expression given in Eq. (\ref{eq: gamma}), which accounts for the $\gamma^{(2)}$ term and the detailed energy dependence of the conductivity.

Our results suggest that the Mott relation becomes increasingly unreliable at temperatures exceeding the critical temperature $T_c$. The linear energy dependence of the surface state density of states in magnetic TIs contributes to the observed discrepancies at higher temperatures, paralleling behaviors observed in graphene systems. In graphene, the Mott relation holds well at low temperatures but exhibits significant deviations at elevated temperatures due to its linear density of states \cite{graphen1, graphen2}.
}

\item{{\it The violation of the Wiedemann-Franz law in magnetic TIs}:

The Wiedemann-Franz law establishes a relationship between the electrical and thermal conductivities of metals, expressed as $\kappa/(\sigma T)= L$, where $\kappa$ is thermal conductivity, and $L$ is the Lorenz number \cite{Mahan, WF-Law}. This law primarily applies to good conductors but is often violated in materials such as topological Kondo model \cite{Buccheri2022}, graphene \cite{graphen2}, and Weyl-semimetal ${\rm WP_2}$ \cite{Jaoui2018}. 

\begin{figure}
	\centering
	\includegraphics[width=0.5\textwidth]{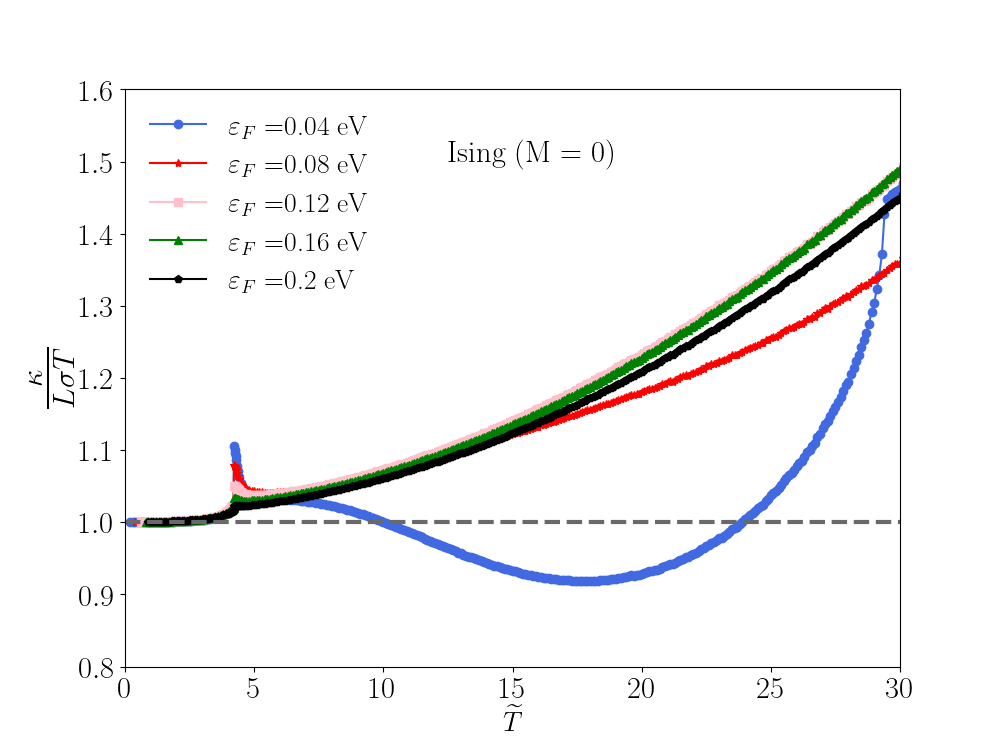}
	\caption{An illustration of the violation of the Wiedemann-Franz law as the temperature \(\widetilde{T} = k_{\rm B} T/(J_1 \tilde{s}^2)\) increases. Various Fermi energies are considered in this analysis. The magnetic atoms on the surface of the TI interact via an Ising Hamiltonian with NN exchange coupling \(J_1\). The interlayer interaction \(J_c\) is scaled to \(J_1\) and set to \(-0.18\). We choose \(\tilde{s} J_1 = 0.3~\text{meV}\), consistent with parameters for manganese bismuth telluride. The dashed line indicates the value predicted by the Wiedemann-Franz law.}
	\label{fig: wiedemann}
\end{figure}
In Fig. \ref{fig: wiedemann} we have plotted the function $\kappa/(L\sigma T)$ versus temperature for different vaues of Fermi energy. As seen at low temperatures, the ratio of thermal to electrical conductivity adheres to the Wiedemann-Franz law. However, the emergence of magnetic clusters near the critical temperature make a violation from this law. As temperature increases, the deviation from the Wiedemann-Franz law becomes increasingly pronounced.

The Wiedemann-Franz law can be derived mathematically using the Sommerfeld expansion, which itself originates from a Taylor expansion of thermodynamic integrals around the Fermi energy. A key integral in this derivation is expressed as:
\begin{equation}
	\int_{-\infty}^{\infty} g(\ep) \left( -\frac{\partial f^0}{\partial \ep} \right) d\ep.
\end{equation}
where \(f^0\) is the Fermi-Dirac distribution function. The accuracy of the Sommerfeld expansion, and consequently the validity of the Wiedemann-Franz law, depends critically on the behavior of the function \(g(\E)\) in the vicinity of the Fermi energy \(\E_{\rm F}\). Specifically, the expansion assumes that \(g(\E)\) can be well approximated by a polynomial expansion around \(\E_{\rm F}\). When \(g(\E)\) varies smoothly and approximately polynomially near \(\E_{\rm F}\), the Sommerfeld expansion provides accurate results, supporting the Wiedemann-Franz law.

However, if the energy dependence of the relaxation time or other transport parameters introduces non-polynomial features into \(g(\E)\), the Taylor expansion truncated to a few terms becomes invalid. This leads to inaccuracies in the Sommerfeld expansion and can result in deviations from the Wiedemann-Franz law. Such deviations are observed experimentally in systems where the scattering mechanisms or band structures induce complex energy dependencies that cannot be captured by low-order polynomial approximations.
}

\item{{\it Validity of Boltzmann transport equation}:
		
		It is important to acknowledge that the Boltzmann transport equation relies on the assumption of weak scattering. However, as we approach the magnetic critical temperature, scattering becomes strong and potentially non-perturbative, which limits the applicability of the relaxation time approximation at the point where the thermopower peaks. Close to the critical temperature, significant scattering occurs, necessitating the inclusion of additional corrections in the $s-d$ Hamiltonian. We have partially addressed this through our scattering potential. While we focus on the average cluster size, we can still utilize the Boltzmann transport equation effectively. However, to achieve greater accuracy, it is essential to consider higher-order effects. Here, we have concentrated on the first-order approximation of the Boltzmann transport equation, leaving higher-order approximations for future investigation.
	}
	
	\item{{\it Intrinsic surface gap in $\rm MnBi_2Te_4$}:
		
Experimental findings on ${\rm MnBi_2Te_4}$ indicate that the nontrivial surface state exhibits a gap of approximately $85~\text{meV}$, even at temperatures significantly exceeding the Neel temperature \cite{experiment_2019}. This intrinsic gap in the surface states is likely attributed to strong spin fluctuations within the material. In this study, we have not considered the impact of this energy gap on the energy spectrum of incident and scattered Dirac electrons. 
While the presence of a gap does not substantially alter the longitudinal resistivity behavior, it does give rise to phenomena such as skew scattering and the associated anomalous Hall effect. 
		
While we have analyzed the mean-field effects of bulk magnetic atoms on the surface electrons, a more comprehensive approach would involve considering the contributions of magnetic atoms in each layer independently, where the distance of each layer from the surface becomes a critical factor \cite{Sun2020}.}

\end{enumerate}

\section*{Acknowledgement}
We would like to express our gratitude to Prof. D. Vashaee for his valuable comments and insightful discussions. This work is based upon research funded by Iran National Science Foundation (INSF) under project No. 4006266.


%\item 
%\item 
%\item 

% Keywords
%\begin{keywords}
% \sep \sep \sep
%\end{keywords}

%\maketitle

% Main text
%\section{}\label{}

% Use the style of numbering in square brackets.
% If nothing is used, default style will be taken.
%\begin{enumerate}[a)]
%\item 
%\item 
%\item 
%\end{enumerate}  

%\begin{itemize}
%\item 
%\item 
%\item 
%\end{itemize}  

%\begin{description}
%\item[]
%\item[] 
%\item[] 
%\end{description}  

%%Remove this from your manuscript

% Figure

%\begin{table}%[]
%\caption{}\label{tbl1}
%\begin{tabular*}{\tblwidth}{@{}LL@{}}
%\toprule
%  &  \\ % Table header row
%\midrule
% & \\
% & \\
% & \\
% & \\
%\bottomrule
%\end{tabular*}
%\end{table}

% Uncomment and use as the case may be
%\begin{theorem} 
%\end{theorem}

% Uncomment and use as the case may be
%\begin{lemma} 
%\end{lemma}


\appendix
\section{Detailed mathematics for determining relaxation times in Eqs. (\ref{eq: tau1}) and (\ref{eq: tau2})}

When the interaction among magnetic ions on the surface of the TI is described by the Ising Hamiltonian, the spins of the magnetic clusters align perpendicular to the surface of the TI. Under these conditions, the $T$-matrix in Eq. (\ref{eq:T1}) is given by:
\begin{equation}
	T^{++}_{\vec{k},\vec{k'}} = \frac{J_0 |s|}{A}\left(\cos^2 \frac{\eta_k}{2} - e^{i\Delta\phi} \sin^2 \frac{\eta_k}{2} \right)F(\xi,\vec{k},\vec{k'}),
\end{equation}
with $\Delta\phi=\phi_{k'} - \phi_{k}$, and
\begin{equation}
F(\xi,\vec{k},\vec{k'})= \int_A d^2r ~e^{ i\vec{\Delta}_{kk'}\cdot \vec{r} - r/\xi},
\label{eq:F-int}
\end{equation}
where $\vec{\Delta}_{kk'}=\vec{k'}-\vec{k}$, $r=|\vec{r}|$, and the integral is performed over the surface $A$ of the TI.

To evaluat the integral in (\ref{eq:F-int}), we employ the Jacobi-Anger expansion:
\begin{equation}
e^{i \vec{\Delta}_{kk'}\cdot\vec{r}} = e^{i \Delta_{kk'} r \cos \alpha} = \sum_{n=-\infty}^{+\infty} i^n J_n(\Delta_{kk'} r) e^{i n \alpha},
\end{equation}
where $\Delta_{kk'}=|\vec{\Delta}_{kk'}|$, and $\alpha$ is the angle between $\vec{r}$ and $\vec{\Delta}_{kk'}$. 

Performing the integral over the angle $\alpha$ yields $\int_0^{2\pi} e^{i \Delta_{kk'} r \cos \alpha} d\alpha = 2\pi J_0(\Delta_{kk'} r)$, where $J_0$ is the Bessel function of the first kind. Using the substitution $r'=r / \xi$, the function $F$ simplifies to:
\begin{align}
	\nonumber F(\xi, \vec{k}, \vec{k'})&= 2\pi \xi^2 \int_0^{\infty} r J_0(\Delta_{kk'} r) e^{-r} dr \\
	%&= 2\pi \xi^2 \beta \sum_{l=0}^{\infty} \frac{(-1)^l}{(l!)^2 2^{2l}} (\Delta \vec{k} \xi)^{2l} \int_0^{\infty} r^{2l+1} e^{-r} \, dr \\
	%&= 2\pi \xi^2 \beta \sum_{l=0}^{\infty} \frac{(-1)^l (2l+1)!}{(l!)^2 2^{2l}} (\Delta \vec{k} \xi)^{2l} \\
	&= \frac{2\pi \xi^2}{\left( 1 + (\Delta_{kk'} \xi)^2 \right)^{3/2}}.
	\label{eq:F2}
\end{align}
Since the scattering is elastic, we have $k = k'$. Using the relation $\Delta_{kk'} = 2k \sin(\Delta \phi/2)$, we obtain the following expression for the scattering amplitude:
\begin{equation}
	\begin{aligned}
		|T^{++}_{\vec{k},\vec{k'}}|^2 &= \frac{4\pi^2 J_0^2 s^2}{A^2} \frac{\xi^4}{[1 + 4k^2 \xi^2\sin^2(\Delta\phi/2)]^3} \\
		&\quad \times (\cos^4 \frac{\eta_k}{2} + \sin^4 \frac{\eta_k}{2}-\frac 12 \sin^2 \eta_k\cos\Delta\phi).
	\end{aligned}
\label{eq:T-matrix}
\end{equation}
As shown in Eq. (\ref{eq:T-matrix}), the scattering amplitude depends only on the angle between the $\vec{k}$ and $\vec{k'}$ vectors. This implies that scattering is isotropic on the surface of the topological insulator, and consequently, the relaxation time depends solely on the magnitude of the wavevector. Therefore, the relaxation time can be calculated using the following standard formula:
\begin{equation}\label{eq: a13}
	\frac{1}{\tau_k}= A \int \frac{d^2 k'}{4\pi^2} \omega(\vec{k}, \vec{k}') (1 - \cos \Delta\phi).
\end{equation}

Using the given integral relations:
\begin{align*}
	\int_0^{2\pi} \frac{1 - \cos \phi}{(a + 1 - a \cos \phi)^3} d\phi &= \frac{(2 + a)\pi}{(1 + 2a)^{5/2}}, \\
	\int_0^{2\pi} \frac{\cos \phi - \cos^2 \phi}{(a + 1 - a \cos \phi)^3} d\phi &= \frac{(a-1)\pi}{(1 + 2a)^{5/2}},
\end{align*}
the final expression for the relaxation time can be directly as:
\begin{equation}
	\frac{1}{\tau_k} = \frac{1}{\tau_0} \bar{n}_c\, k\xi \left( \frac{\xi}{a} \right)^3 \frac{3 + (1 + 4k^2 \xi^2) \cos^2 \eta_k}{\sin \eta_k \left( 1 + 4k^2 \xi^2 \right)^{5/2}}.
\end{equation}
%where $\tau_0 = \frac{\sqrt{3} \hbar^2 v_{\rm F}}{2\pi^2 J_0^2 s^2 a}$ is a constant with units of time, and $\bar{n}_c = n_c / N$ denotes the average number of magnetic clusters per triangular lattice site, with $N$ being the total number of sites.

%%%%%%%%%%%%%%
%%%%%%%%%%%%%%

To derive Eq. \ref{eq: tau2}, we perform the following steps. 
For a cluster with spin oriented at an arbitrary direction given by the angle $(\theta_s, \phi_s)$, the electron-cluster interaction is given by:
\begin{align}
\vec{\sigma} \cdot \vec{s} = s
\begin{pmatrix}
	\cos \theta_s &  \sin \theta_s e^{-i\phi_s}\\
	\sin \theta_s e^{i\phi_s} & -\cos \theta_s
\end{pmatrix}.
\end{align}
The scattering amplitude is then given by:
\begin{align}
|T^{++}_{\vec{k},\vec{k'}}|^2& = \frac{4\pi^2 J_0^2 s^2}{A^2}\frac{\xi^4_{\theta_s} \Gamma(\theta_s,\phi_s;\eta_k,\Delta\phi)}{[1 + 4k^2 \xi^2_{\theta_s}\sin^2(\Delta\phi/2)]^3},
\end{align}
where the function $\Gamma$ is
\begin{align}
	\Gamma &= \cos^2\theta_s[1-\sin^2\eta_k\sin^2(\Delta\phi/2)] \nonumber \\
	&+\sin \eta_k \sin[\frac{\phi + \phi'}{2}-\phi_s] \times \nonumber \\
	&(\sin \eta_k \sin^2\theta_s \sin [\frac{\phi + \phi'}{2}-\phi_s]-\cos \eta_k \cos \frac{\Delta\phi}{2} \sin 2\theta_s).\nonumber
\end{align}
As observed, in this case, the scattering amplitude depends not only on $\Delta \phi$, but also on $\phi_k$ and $\phi_{k'}$. This suggests that, in the presence of Heisenberg interactions among magnetic atoms on the surface of the TI, the surface exhibits anisotropic behavior. However, as our simulations show, the combined effects of the isotropic Heisenberg interaction and the on-site anisotropy term in Eq. (\ref{eq:H1}) lead to symmetric spin configurations. In these configurations, clusters tend to align such that their in-plane spin components cancel out on average, as illustrated in Fig. \ref{fig: clustering-schematic}. Specifically, for a cluster oriented at $(\theta_s, \phi_s)$, there exists a corresponding cluster at $(\theta_s, -\phi_s)$, resulting in a net spin component primarily along the $z$-axis. To incorporate these symmetries, we perform an integral over $\Gamma$ with respect to $\phi_s$ as:
\begin{align}
	\Gamma(\theta_s;\eta_k,\Delta\phi) 
	\nonumber&= \frac{1}{2\pi} \int_{0}^{2\pi} \Gamma(\theta_s,\phi_s;\eta_k,\Delta\phi) d\phi_s \\
	\nonumber&=\cos^2\theta_s[1-\sin^2\eta_k\sin^2(\Delta\phi/2)] \\
	&+ \frac{1}{2} \sin^2\theta_s\sin^2\eta_k .
\label{eq:Gamma-Ave}
\end{align}
Using Eq. (\ref{eq:Gamma-Ave}), the scattering amplitude of surface Dirac electrons from clusters with spin \( s_z = s \cos \theta_s \), averaged over the azimuthal angle \( \phi_s \), is given by:
\begin{align}
\langle|T^{++}_{\vec{k},\vec{k'}}|^2\rangle_{\phi_s}& = \frac{4\pi^2 J_0^2 s^2}{A^2}\frac{\xi^4_{\theta_s}\Gamma(\theta_s;\eta_k,\Delta\phi)}{[1 + 4k^2 \xi^2_{\theta_s}\sin^2(\Delta\phi/2)]^3}.
\label{eq:T-Ave}
\end{align}
Using Eqs. (\ref{eq: a13}) and (\ref{eq:T-Ave}), the relaxation time becomes:
\begin{align}
	\frac{1}{\tau_{k,\theta_s}} &= \frac{4\pi J_0^2 s^2}{A\hbar} n_{\theta_s}\xi^4_{\theta_s}\times\nonumber\\
& \int d^2k' \frac{\sin^2(\Delta\phi/2)\Gamma(\theta_s;\eta_k,\Delta\phi)}{[1 + 4k^2 \xi^2_{\theta_s}\sin^2(\Delta\phi/2)]^3}\delta(\E_k - \E_{k'}).
\end{align}
Given that for a triangular lattice with lattice spacing $a$, the total area is $A = N^2 (\sqrt{3} a^2/2)$, and defining the normalized density as $\bar{n}_{\theta_s} = n_{\theta_s} / N$, evaluating this integral leads to the expression for the relaxation time presented in Eq. (\ref{eq: tau2}).

%\section{}\label{}


%\bibliographystyle{cas-model2-names}

%\bibliographystyle{plain}
%\bibliographystyle{unsrt} % Change the bibliography style to "unsrt"
%\bibliography{cas-refs.bib}

\begin{thebibliography}{99}
	\bibitem{Fenwick2020}
	O. Fenwick, and A. Jones, Materials for the Energy Transition roadmap: Materials Thermoelectric Energy Conversion Materials, Henry Royce Institute (2020).
	
	\bibitem{Gayner2016}
	Chhatrasal Gayner, Kamal K. Kar, Recent advances in thermoelectric materials, Progress in Materials Science {\bf 83}, 330 (2016).
	
	\bibitem{Singh2024}
	Rakesh Singh, Surya Dogra, Saurav Dixit, Nikolai Ivanovich Vatin, Rajesh Bhardwaj, Ashok K. Sundramoorthy, H.C.S. Perera, Shashikant P. Patole, Rajneesh Kumar Mishra, Sandeep Arya, Advancements in thermoelectric materials for efficient waste heat recovery and renewable energy generation, Hybrid Advances {\bf 5}, 100176 (2024).
	
	\bibitem{Ivanov2018}
	Yuri V. Ivanov, Alexander T. Burkov, and Dmitry A. Pshenay-Severin, Thermoelectric Properties of Topological Insulators, Phys. Status Solidi B, 1800020 (2018).
	
	\bibitem{Zheng2019}
	Y. Zheng , T. Lu, Md M. H. Polash, M. Rasoulianboroujeni, N. Liu, M. E. Manley, Y. Deng, P. J. Sun, X. L. Chen, R. P. Hermann, D. Vashaee, J. P. Heremans, and H. Zhao, Paramagnon drag in high thermoelectric figure of merit Li-doped ${\rm MnTe}$, Sci. Adv. {\bf 5}, eaat9461 (2019).
	
	\bibitem{Polash2020}
	M. M. H. Polash, F. Mohaddes, M. Rasoulianboroujeni, and D. Vashaee, Magnon-drag thermopower in antiferromagnets versus ferromagnets, J. Mater. Chem. C {\bf 8}, 4049 (2020).
	
	\bibitem{Polash2021}
	M. M. H. Polash, D. Moseley, J. Zhang, R. P. Hermann, and D. Vashaee, Understanding and design of spin-driven thermoelectrics, Cell Rep. Phys. Sci. {\bf 2}, 100614 (2021).
	
	\bibitem{Heydarinasab2024}
	F. Heydarinasab, M. Jazandari, M. M. H. Polash, J. Abouie, and D. Vashaee, Paramagnon heat capacity and anomalous thermopower in anisotropic magnetic systems: Understanding interlayer spin correlations in a magnetically disordered phase, Phys. Rev. B {\bf 109}, 054418 (2024).
	
	\bibitem{Xu2017}
	N. Xu, Y. Xu, and J. Zhu, Topological insulators for thermoelectrics, npj Quant Mater {\bf 2}, 51 (2017).
	
	\bibitem{Pan2025}
	Yu Pan, Bin He, Xiaolong Feng, Fan Li, Dong Chen, Ulrich Burkhardt, and Claudia Felser,
	A magneto-thermoelectric with a high figure of merit in topological insulator $\rm Bi_{88}Sb_{12}$, Nat. Mater. {\bf 24}, 76 (2025).
	
	\bibitem{Moore2010}
	Joel E. Moore, The birth of topological insulators, nature {\bf 464}, 194 (2010).
	
	\bibitem{Tokura2019}
	Yoshinori Tokura, Kenji Yasuda, and Atsushi Tsukazaki, Magnetic topological insulators, Nat. Rev. Phys. {\bf 1}, 126 (2019). 
	
	\bibitem{Vyborny2009}
	K. Vyborny, A. A. Kovalev, J. Sinova, and T. Jungwirth, Semiclassical framework for the calculation of transport anisotropies, Phys. Rev. B {\bf 79}, 045427 (2009).
	
	\bibitem{Sabzali2015}
	A. Sabzalipour, J. Abouie, and S. H. Abedinpour, Anisotropic conductivity in magnetic topological insulators, J. Phys.: Condens. Matter {\bf 27}, 115301 (2015).
	
	\bibitem{Zarezad2018}
	A. N. Zarezad, and J. Abouie, Transport in magnetically doped topological insulators: Effects of magnetic clusters, Phys. Rev. B {\bf 98}, 155413 (2018).
	
	\bibitem{Zarezad2020}
	A. N. Zarezad, and J. Abouie, Transport in two-dimensional Rashba electron systems doped with interacting magnetic impurities, Phys. Rev. B {\bf 101}, 115412 (2020).
	
	\bibitem{diep2011}
	K. Akabli, Y. Magnin, Masataka Oko, Isao Harada, and H. T. Diep, Phys. Rev. B {\bf 84}, 024428 (2011).
	
	\bibitem{review_onMTI2021}
	Pinyuan Wang, Jun Ge, Jiaheng Li, Yanzhao Liu, Yong Xu, Jian Wang, Intrinsic magnetic topological insulators, Innov. {\bf 2}, 2-100098 (2021).
	
	\bibitem{PRL_MagneticInteractions}
	Bing Li, J. -Q. Yan, D. M. Pajerowski, Elijah Gordon, A. -M. Nedic, Y. Sizyuk, Liqin Ke, P. P. Orth, D. Vaknin, and R. J. McQueeney,
	Competing magnetic interactions in the antiferromagnetic topological insulator \(\rm MnBi_{2}Te_{4}\), Phys. Rev. Lett. {\bf 124}, 167204 (2020).
	
	\bibitem{Li2021}
	Bing Li, D. M. Pajerowski, S. X. M. Riberolles, Liqin Ke, J. -Q. Yan, and R. J. McQueeney, 
	Quasi-two-dimensional ferromagnetism and anisotropic interlayer couplings in the magnetic topological insulator \(\rm MnBi_{2}Te_{4}\), Phys. Rev. B {\bf 104}, L220402 (2021).
	
	\bibitem{GMR1}
	A. Fert, and I. A. Campbell, Two-current conduction in nickel, Phys. Rev. Lett. {\bf 21}, 1190 (1968).
	
	\bibitem{GMR2}
	M. N. Baibich, J. M. Broto, A. Fert, F. Nguyen Van Dau, and F. Petroff, Giant magnetoresistance of ${\rm (001)Fe/(001)Cr}$ magnetic superlattices, Phys. Rev. Lett. {\bf 61}, 2472 (1988).
	
	\bibitem{phase1}
	Alla E. Petrova, E. D. Bauer, Vladimir Krasnorussky, and Sergei M. Stishov, Behavior of the electrical resistivity of MnSi at the ferromagnetic phase transition, Phys. Rev. B {\bf 74}, 092401 (2006).
	
	\bibitem{phase2}
	S. M. Stishov, A. E. Petrova, S. Khasanov, G. Kh. Panova, A. A. Shikov, J. C. Lashley, D. Wu, and T. A. Lograsso, Magnetic phase transition in the itinerant helimagnet MnSi: Thermodynamic and transport properties, Phys. Rev. B {\bf 76}, 052405 (2007).
	
	\bibitem{Zhi-Huan2009}
	Luo Zhi-Huan, Loan Mushtaq, Liu Yan and Lin Jian-Rong, Critical behaviour of the ferromagnetic Ising model on a triangular lattice, Chinese Phys. B {\bf 18}, 2696 (2009).
	
	\bibitem{Matsubara1996}
	K. Murao, F. Matsubara, and T. Kudo, Anisotropic ferromagnet on a triangular lattice with antiferromagnetic next-nearest-neighbor interactions, J. Phys. Soc. Jpn., {\bf 65} 1399 (1996).
	
	\bibitem{wolff1}
	U. Wolff, Lattice field theory as a percolation process, Phys. Rev. Lett. {\bf 60}, 1461 (1988).
	
	\bibitem{Coniglio}
	Coniglio, A and Klein, Clusters and Ising critical droplets: a renormalisation group approach, Journal of Physics A: Mathematical and General {\bf 13}, 2775 (1980).
	
	\bibitem{hoshen}
	J. Hoshen and R. Kopelman, Percolation and cluster distribution. I. Cluster multiple labeling technique and critical concentration algorithm, Phys. Rev. B {\bf 14}, 3438 (1976).
	
	\bibitem{Otrokov2019}
	M. M. Otrokov, I. P. Rusinov, M. Blanco-Rey, M. Hoffmann, A. Yu. Vyazovskaya, S. V. Eremeev, A. Ernst, P. M. Echenique, A. Arnau, E. V. Chulkov,
	Unique thickness-dependent properties of the van der Waals interlayer antiferromagnet ${\rm MnBi_2Te_4}$ films, Phys. Rev. Lett. {\bf 122}, 107202 (2019).
	
	\bibitem{Zhao2021}
	Yi-Fan Zhao, Ling-Jie Zhou, Fei Wang, Guang Wang, Tiancheng Song, Dmitry Ovchinnikov, Hemian Yi, Ruobing Mei, Ke Wang, Moses H W Chan, Chao-Xing Liu, Xiaodong Xu, Cui-Zu Chang, Even-Odd layer-dependent anomalous Hall effect in topological magnet ${\rm MnBi_2Te_4}$ thin films, Nano. Lett. {\bf 21}, 7691 (2021).
	
	\bibitem{Shao2021}
	Jifeng Shao, Yuntian Liu, Meng Zeng, Jingyuan Li, Xuefeng Wu, Xiao-Ming Ma, Feng Jin, Ruie Lu, Yichen Sun, Mingqiang Gu, Kedong Wang, Wenbin Wu, Liusuo Wu, Chang Liu, Qihang Liu, and Yue Zhao, Pressure-tuned intralayer exchange in superlattice-like \(\rm MnBi_2Te_4/(Bi_2Te_3)_n\) topological insulators, Nano Lett. {\bf 21}, 5874 (2021).
	
	\bibitem{J0_2022}
	Padmanabhan H, Stoica VA, Kim PK, Poore M, Yang T, Shen X, Reid AH, Lin MF, Park S, Yang J, Wang HH, Koocher NZ, Puggioni D, Georgescu AB, Min L, Lee SH, Mao Z, Rondinelli JM, Lindenberg AM, Chen LQ, Wang X, Averitt RD, Freeland JW, Gopalan V., Large Exchange Coupling Between Localized Spins and Topological Bands in ${\rm MnBi_2Te_4}$, Adv Mater. {\bf 49}, e2202841 (2022).
	
	\bibitem{flakes1}
	Dong Sun Lee, Tae-Hoon Kim, Cheol-Hee Park, Chan-Yeup Chung, Young Soo Lim, Won-Seon Seoa and Hyung-Ho Park,   
	Crystal structure, properties and nanostructuring of a new layered chalcogenide semiconductor, \(\rm MnBi_2Te_4\), Cryst. Eng. Comm. {\bf 15}, 5532–5538 (2013).
	
	\bibitem{flakes2}
	Yun Zheng, Xian Yi Tan, Xiaojuan Wan, Xin Cheng, Zhihong Liu, Qingyu Yan, 
	Thermal stability and mechanical response of \(\rm Bi_{2}Te_{3}\)-based materials for thermoelectric applications, Applied Energy Materials {\bf 3}, 2078 (2020).
	
	\bibitem{Htemperature1}
	Le Fang, Chen Chen, Xionggang Lu, and Wei Ren,
	Effects of pressure and temperature on topological electronic materials \(\rm X_2Y_3 (X=As, Sb, Bi; Y=Se, Te)\) using first-principles, Phys. Chem. Chem. Phys. {\bf 25}, 20969 (2023).
	
	\bibitem{Htemperature2}
	Jonas Anversa, Sudip Chakraborty, Paulo Piquini, and Rajeev Ahuja,
	High pressure driven superconducting critical temperature tuning in \(\rm Sb_{2}Se_{3}\) topological insulator, Appl. Phys. Lett. {\bf 108}, 212601 (2016).
	
	\bibitem{Tang2015}
	Zhenglong Tang, Lipeng Hu, Tiejun Zhu, Xiaohua Liu, and Xinbing Zhao,
	High performance $n$-type Bismuth Telluride based alloys for mid-temperature power generation, J. Mater. Chem. C {\bf 3}, 10597 (2015). 
	
	\bibitem{experiment_2019}
	Seng Huat Lee, Yanglin Zhu, Yu Wang, Leixin Miao, Timothy Pillsbury, Hemian Yi, Susan Kempinger, Jin Hu, Colin A. Heikes, P. Quarterman, William Ratcliff, Julie A. Borchers, Heda Zhang, Xianglin Ke, David Graf, Nasim Alem, Cui-Zu Chang, Nitin Samarth, and Zhiqiang Mao, Spin scattering and noncollinear spin structure-induced intrinsic anomalous Hall effect in antiferromagnetic topological insulator \(\rm MnBi_{2}Te_{4}\), Phys. Rev. Research {\bf 1}, 012011(R) (2019).
	
	\bibitem{experiment_2020}
	Hao Li, Shengsheng Liu, Chang Liu, Jinsong Zhang, Yong Xu, Rong Yu, Yang Wu, Yuegang Zhang, and Shoushan Fan,
	Antiferromagnetic topological insulator ${\rm MnBi_2Te_4}$: synthesis and magnetic properties, Phys. Chem. Chem. Phys. {\bf 22}, 556 (2020).

\bibitem{Jazandari2025}
M. Jazandari, J. Abouie, and D. Vashaee, What Really Drives Thermopower: Specific Heat or Entropy as the Unifying Principle Across Magnetic, Superconducting, and Nanoscale Systems, arXiv:2506.06745.

\bibitem{Sugihara1972}
K. Sugihara, Magnon Drag Effect in Magnetic Semiconductors, J. Phys. Chem. Solids 33, 1365 (1972).

\bibitem{Zanmarchi1968}
G. Zanmarchi and C. Haas, Magnon Drag at Optical Frequencies and the Infrared Spectrum of MnTe, J. Appl. Phys. 39, 596 (1968).

\bibitem{Ko2022}
Wonhee Ko, Saban Hus, Hoil Kim, Jun Sung Kim, Xiao-Guang Zhang, and An-Ping Li, Resistivity of Surface Steps in Bulk-Insulating Topological Insulators, Front. Mater. {\bf 9}, 887484 (2022)
	
	\bibitem{Yan2019}
	J.-Q. Yan, Q. Zhang, T. Heitmann, Zengle Huang, K. Y. Chen, J.-G. Cheng, Weida Wu, D. Vaknin, B. C. Sales, and R. J. McQueeney,
	Crystal growth and magnetic structure of ${\rm MnBi_2Te_4}$, Phys. Rev. Materials {\bf 3}, 064202 (2019).
	
	\bibitem{experiment_2022}
	H. Zhang, C. Q. Xu, S. H. Lee, Z. Q. Mao, and X. Ke, Thermal and thermoelectric properties of an antiferromagnetic topological insulator \(\rm MnBi_{2}Te_{4}\), Phys. Rev. B {\bf 105}, 184411 (2022).
	
	\bibitem{Mahan}
	M. Jonson, and G. D. Mahan, Mott's formula for the thermopower and the Wiedemann-Franz law, Phys. Rev. B {\bf 21}, 4223 (1980).
	
	\bibitem{graphen1}
	Tomas Löfwander and Mikael Fogelström, Impurity scattering and Mott’s formula in graphene, Phys. Rev. B {\bf 76}, 193401 (2007).
	
	\bibitem{graphen2}
	P. Ray, K. Sarkar, Mott and Wiedemann Franz Law for Monolayer Graphene for Different Scattering Mechanisms. In: Mishra, Y.K., Lingamallu, G., Ghosh, T. (eds) Selected Articles from the 2nd International Conference on Advanced Nanomaterials and Applications. ICANA 2024. Springer Proceedings in Physics, vol 413. Springer, Singapore, (2025).
	
%	\bibitem{dots}
%	M. A. Manya , G. B. Martins, and M. S. Figueira, Spin-orbit coupling effects on thermoelectric transport properties in quantum dots, Phys. Rev. B {\bf 105}, 165421 (2022).
	
	\bibitem{WF-Law}
	A. Grunwald and J. Hajdu, Note on the Wiedemann-Franz law and the Mott rule in two-dimensional systems in strong magnetic fields, Z. Phys. B Condensed Matter {\bf 60}, 235 (1985).
	
	\bibitem{Buccheri2022}
	Francesco Buccheri, Andrea Nava, Reinhold Egger, Pasquale Sodano, and Domenico Giuliano,
	Violation of the Wiedemann-Franz law in the topological Kondo model, Phys. Rev. B {\bf 105}, L081403 (2022).
	
	\bibitem{Jaoui2018}
	Alexandre Jaoui, Benoit Fauque, Carl Willem Rischau, Alaska Subedi, Chenguang Fu, Johannes Gooth, Nitesh Kumar, Vicky Süß, Dmitrii L. Maslov, Claudia Felser, and Kamran Behnia, Departure from the Wiedemann-Franz law in WP2 driven by mismatch in T-square resistivity prefactors, npj Quant Mater {\bf 3}, 64 (2018).
	
	
	\bibitem{Sun2020}
	Hai-Peng Sun, C. M. Wang, Song-Bo Zhang, Rui Chen, Yue Zhao, Chang Liu, Qihang Liu, Chaoyu Chen, Hai-Zhou Lu, and X. C. Xie, Analytical solution for the surface states of the antiferromagnetic topological insulator ${\rm MnBi_2Te_4}$, Phys. Rev. B {\bf 102}, 241406(R) (2020).

	
	
\end{thebibliography}



\end{document}